\documentclass[acmsmall,screen]{acmart}

\AtBeginDocument{%
  \providecommand\BibTeX{{%
    \normalfont B\kern-0.5em{\scshape i\kern-0.25em b}\kern-0.8em\TeX}}}

\acmBooktitle{CSUR '21: ACM Computing Survey}

\usepackage{lipsum}
\usepackage{tcolorbox}
\usepackage{longtable}
\usepackage[inline]{enumitem}
\usepackage{multirow}

\definecolor{codegreen}{rgb}{0,0.6,0}
\definecolor{codegray}{rgb}{0.5,0.5,0.5}
\definecolor{codepurple}{rgb}{0.58,0,0.82}
\definecolor{backcolour}{rgb}{0.95,0.95,0.92}

\definecolor{delim}{RGB}{20,105,176}
\definecolor{numb}{RGB}{106, 109, 32}
\definecolor{string}{rgb}{0.64,0.08,0.08}

\usepackage{listings}
\begin{document}

\title{What are the attackers doing now? Automating cyberthreat intelligence extraction from text on pace with the changing threat landscape: A survey}

\author{Md Rayhanur Rahman}
\email{mrahman@ncsu.edu}
\author{Rezvan Mahdavi-Hezaveh}
\email{rmahdav@ncsu.edu}
\author{Laurie Williams}
\email{lawilli3@ncsu.edu}
\affiliation{
  \institution{North Carolina State University}
  \city{Raleigh}
  \state{North Carolina}
  \country{USA}
}

\renewcommand{\shortauthors}{Rahman et al.}

\begin{abstract}

Cyberattackers are continuously changing their strategies and techniques to bypass the security mechanisms deployed by the targeted organizations. To thwart these attackers, cyberthreat intelligence (CTI) can help organizations keep pace with ever-changing threat landscapes. Cybersecurity researchers have contributed to the automated extraction of CTI from textual sources, such as threat reports and online articles, where cyberattack strategies, procedures, and tools are described. 
\textit{The goal of this article is to aid cybersecurity researchers understand the current techniques used for cyberthreat intelligence extraction from text through a survey of relevant studies in the literature.} 
We systematically collect "CTI extraction from text"-related studies from the literature and categorize the CTI extraction purposes. We propose a CTI extraction pipeline abstracted from these studies. We identify the data sources,  techniques, and CTI sharing formats utilized in the context of the proposed pipeline. Our work finds ten types of extraction purposes, such as extraction indicators of compromise extraction, TTPs (tactics, techniques, procedures of attack), and cybersecurity keywords. We also identify seven types of textual sources for CTI extraction, and textual data obtained from hacker forums, threat reports, social media posts, and online news articles have been used by almost 90\% of the studies. Natural language processing along with both supervised and unsupervised machine learning techniques such as named entity recognition, topic modelling, dependency parsing, supervised classification, and clustering are used for CTI extraction. We observe the technical challenges associated with these studies related to obtaining available clean, labelled data which could assure replication, validation, and further extension of the studies. As we find the studies focusing on CTI information extraction from text, we advocate for building upon the current CTI extraction work to help cybersecurity practitioners with proactive decision making such as threat prioritization, automated threat modelling to utilize knowledge from past cybersecurity incidents. 
\end{abstract}


\keywords{Cyberthreat Intelligence, CTI extraction, CTI Mining, IoC extraction, TTPs extraction, attack pattern extraction, threat reports, tactical threat intelligence, technical threat intelligence}

\maketitle

\section{Introduction}
\label{sec:introduction}
Defending and preventing cyberattacks have become increasingly difficult as attack tactics and techniques are continuously evolving. The attackers' community is now more organized in its operation and is motivated by financial reasons~\cite{hackernews-financial-backup} as well. Cyberattack trends now have shifted from small group efforts to larger organized crime. For example, in 2020, the University of Utah experienced a ransomware attack where sensitive student information was stolen by an attacker group and the university suffered a financial loss of \$457,000~\cite{uatutahransomware}.  

To keep pace with attackers' ever-changing ways of launching cyberattacks, cyberthreat intelligence (CTI), also known as ``threat intelligence'' can be used to help information technology (IT) organizations proactively defend against cyberattacks. According to the definition provided by McMillan et al.~\cite{mcmillan2013definition}, ``Threat Intelligence is evidence-based knowledge, including context, mechanisms, indicators, implications, and action-oriented advice about an existing or emerging menace or hazard to assets. This intelligence can be used to inform decisions regarding the subject’s response to that menace or hazard.'' Hence, CTI is a set of organized collected information about cyberthreats that can be utilized to predict, prevent, or defend against cyberattacks \cite{tounsi2018survey, legoy2019retrieving}. CTI can help IT organizations build the necessary tactics and strategies to weaken the attacker’s methods as well as to build tools and techniques to thwart malicious attempts. 
For example, in 2020, Amazon revealed that the company's AWS Shield, developed by the company's threat analysis group and security team, defended a state-sponsored distributed denial of service (DDoS) attack with a peak traffic volume of 2.3 Tbps, which is the largest DDoS attack to date~\cite{awsddos}.  

CTI can be extracted, aggregated, synthesized, and analyzed from publicly available cyberthreat-related documents, articles, reports, social media, and human intelligence~\cite{ctidefwiki}. These information sources can contain how the attackers target the organization, what their strategies are, which tools and procedures are utilized, and detailed descriptions of how the attacks are performed. As the number and types of attacks have grown, so has the volume of textual content focusing on cyberthreat news, attack patterns, tools, and techniques. The attack tactics and techniques are also consistently evolving.  Extracting the most important information has become a challenging task due to the large volume of data, noise, anomalies, and difficulty in establishing correlation among the obtained information. Moreover, these data are in textual formats written in natural language (i.e., English). Hence, manually extracting relevant information from the aforementioned CTI information sources can be error-prone and inefficient~\cite{zhao2020timiner}.

Cybersecurity researchers and practitioners have focused on the automatic extraction of CTI information, mainly utilizing natural language processing (NLP) and machine learning (ML) techniques ~\cite{liao2016acing, zhu2018chainsmith, lee2017sec}. Studies on extracting malware indicators, attack patterns, and generating cyberthreat alerts can be found in the literature. A systematization of these studies, categorizing the CTI extraction purposes and identifying associated techniques facilitate the extension and improvement of current work and introduces future research paths in the CTI extraction domain.   

\textit{The goal of this article is to aid cybersecurity researchers understand the current techniques used for cyberthreat intelligence extraction from text through a survey of relevant studies in the literature.} To achieve this goal, using a keyword search, we collect existing studies on automatic extraction of CTI from textual descriptions. We use open coding \cite{saldana2015coding} and card sorting \cite{spencer2004card} techniques to perform a qualitative analysis of these studies. We list our contributions as following: 

\begin{itemize}[leftmargin=3mm]
    \item A systematic categorization of the CTI extraction purposes (such as extracting malware indicators, attack patterns from the text) performed in these studies\footnote{throughout this article \textit{studies} refers to the set of  studies related to CTI extraction from text in this survey} (Section~\ref{goal}),
    
    \item CTI extraction pipeline, where the pipeline abstracts the steps for CTI extraction observed in the studies (Section~\ref{pipeline}),
    
    \item A categorization of  NLP and ML techniques associated with CTI extraction (Section~\ref{mining-approach}),
    
    \item A set of textual data sources and CTI sharing formats utilized in the studies for CTI extraction and sharing, respectively (Section~\ref{data-source} and \ref{sharing-format}),
    
    \item A compilation of application scenarios of the extracted CTI demonstrated by the researchers in the studies (Section~\ref{use-cases}).
    
    \item A set of recommendations to facilitate cybersecurity researchers to conduct future research in the CTI extraction domain (Section~\ref{future}).
\end{itemize}

The rest of the article is organized as follows. In Section~\ref{definition}, we discuss CTI and its role in proactive defense. In Section~\ref{related-work}, we discuss the related survey studies. In Section~\ref{methodlogy}, we discuss our methodology. In Section~\ref{goal}, we discuss the types of CTI extraction purpose observed in the studies. In Section~\ref{pipeline}, we discuss the CTI extraction pipeline along with each step in the pipeline. In Section~\ref{challenge} and ~\ref{future}, we discuss several insights for further research work. Finally, in Section~\ref{discussion}, we conclude the article.  

\section{Cyberthreat intelligence}
\label{definition}
Cybersecurity is a balancing act between cyber-attackers and targeted entities/organizations by attackers~\cite{schneier1998security}. The attackers are constantly probing to exploit security weaknesses in the system, while the organizations constantly monitor and defend the attack attempts. The way the modern IT infrastructure currently operates can be susceptible to security weaknesses, such as insecure coding practices, zero-day exploits, inconsistent patching, vulnerable third-party libraries, data exposure, human error, and social engineering. These factors can provide the attackers advantages in launching cyberattacks. Thus, handling cyberthreats in a \textit{reactive} manner, such as only responding to security incidents, can make the organization more vulnerable to cyberthreats. To address this issue, organizations should be \textit{proactive} about defending against attackers who are always on the move in an ever-changing threat landscape.

\subsection{Definition}
\label{cti-def}
Cyberthreat intelligence (CTI), also known as threat intelligence, can be used as a proactive defense mechanism against cyberattacks. McMillan et al. \cite{mcmillan2013definition} define CTI as ``evidence-based knowledge, including context, mechanisms, indicators, implications, and actionable advice about an existing or emerging menace or hazard to assets that can be used to inform decisions regarding the subject's response to that menace or hazard.'' Dalziel et al. define CTI as refined, analyzed, or processed data that should be at least potentially relevant to the organization and/or objectives, specific enough to prompt some response, action, or decision, and contribute to any useful business outcome \cite{dalziel2014define}. CTI\footnote{Although we provide several definitions of CTI, there is no universally accepted definition of CTI~\cite{tounsi2018survey, Abu2018Cyberthreat, sauerwein2017threat}} can also be used for predicting, preventing, or defending an attack, shortening the window between compromise and detection, and helping to clarify the risk landscape \cite{tounsi2018survey, mcmillan2013definition, legoy2019retrieving}. 

\subsection{CTI subdomains} \label{def:categories}
According to Tounsi et al. \cite{tounsi2018survey}, CTI can be categorized into four subdomains  as follows:

\begin{enumerate}[leftmargin=5mm]
    \item \textbf{Strategic CTI} which is the information generally in the form of reports, briefings, or conversations that help security decision-makers understand and identify current and future risks. This information includes financial effects, trends, and historical data of cyberattacks. This kind of CTI is targeted for the nontechnical audience~\cite{crowdstrike-cti-def}.
    
    \item \textbf{Operational CTI} which is information about specific impending attacks on an organization. This kind of CTI helps organizations understand the relevant factors of specific attacks, such as nature, timing, intent, and threat actors. This kind of CTI is targeted towards security professionals working in security operation centers~\cite{crowdstrike-cti-def}. 
    
    \item \textbf{Tactical CTI} which is often referred to as Tactics, Techniques, and Procedures (TTPs)~\cite{tounsi2018survey, ti-dns-stuff, attack}. A \textit{tactic} is the highest-level description of a cyberthreat - the goal of the adversary (``why''). A \textit{technique} gives a more detailed description of cyberthreat in the context of a tactic - ``how'' to achieve an objective and ``what'' is gained. \textit{Procedures} are even lower-level, highly detailed descriptions (containing tools and attack group description) in the context of a technique. The tactical CTI is used by security experts to improve their defense strategies for current tactics. Technical press, white papers, and threat reports are sources for tactical CTI.
    
    \item \textbf{Technical CTI} which is the set of information usually being consumed through computational resources. For example, information (such as IP address, packet contents) collected by firewalls regarding DDoS attacks can be consumed by analytical and visualization tools. Technical CTI can be collected as indicators of compromise (IoC) referring to the ``forensic evidence of potential intrusions on a host system or network''~\cite{ioc-def-trend-micro} which includes malicious IP address, URLs, malware signatures, anomalous network traffic. Technical CTI usually feeds information towards investigative and monitoring activities inside an organization. 
\end{enumerate}

\subsection{Uses of CTI}
Advantages of utilizing CTI are numerous, which are discussed below.

\begin{itemize}[leftmargin=3mm]
    \item \textbf{Proactive and actionable defense:} In proactive defense, future threat strategies are foreseen and these insights are incorporated into the defense mechanisms of the system \cite{colbaugh2011proactive}. Identifying CTI from previous threats, analysing the identified CTI information, and deriving actionable insights are helpful keys to prepare a system for proactive defense. Moreover, modern day cyberattacks can use multiple means to get propagated and can be active in multiple stages. For example, attackers can infiltrate the network first, spread laterally across all devices, and then compromise systems through vertical propagation. These attacks are armed with zero-day exploits, social engineering, and malicious client side scripts, which help the attackers evade the traditional cyberattack detection and cyber defense mechanism~\cite{tounsi2018survey}. The use of CTI can help organizations to prevent these attacks. 
    
    \item \textbf{Constructing threat profiles:} CTI can be used to build threat profiles for well-known cyberattack groups, which can aid organizations in securing their defense mechanisms based on the attack tactics and techniques from the threat profile.
    
    
    \item \textbf{Knowledge sharing and awareness building:} The collective sharing and exchanging of knowledge gained from CTI information can accelerate the learning and awareness among organizations to prevent cyberattacks. For example, if organizations find an intruder in the active phase of an attack, there are greater chances of defending the attack through collective sharing~\cite{zurkus2015threat}. Moreover, CTI sharing is a cost-effective tool for thwarting cybercrime~\cite{ponemon2014exchanging}.
    
    \item \textbf{Cybersecurity research:} CTI related information, such as attack indicators and TTPs, can be utilized by cybersecurity researchers to develop new insights on cyberthreat domains. 
\end{itemize}

\section{Related studies}
\label{related-work}

In this section, we discuss related literature review or survey studies on different aspects of CTI. 

\textbf{Technical CTI:} Tounsi et al. \cite{tounsi2018survey} first explained the four categories of CTI in the literature that we mention in Section \ref{def:categories}. Their survey focused on technical CTI, existing issues, emerging research, and trends related to technical CTI. One of the issues they found is that identifying related technical CTI is challenging for organizations because of the large amount of available technical CTI. They also evaluated the most popular and available CTI gathering tools and compared their features. 

\textbf{CTI sharing:} Sauerwein et al. \cite{sauerwein2017threat} conducted a study of 22 CTI-sharing platforms that enable automation of the generation, refinement, and examination of security data. Their study resulted in eight key findings including: ``There is no common definition of threat intelligence sharing platforms'', and ``Most platforms focus on data collection instead of analysis''. Wagner et al. \cite{wagner2019cyber} explored the current state-of-the-art approaches and technical and nontechnical challenges in sharing CTI. They used articles from academic and gray literature. Their findings show widely discussed topics in CTI sharing, such as establishing a successful collaboration between stakeholders and automating parts of the CTI sharing process. 

\textbf{Threat analysis:} Tuma et al. \cite{tuma2018threat} performed a systematic literature review on 26 methodologies of cyberthreat analysis. They compared methodologies based on different aspects such as applicability, characteristics of the analysis procedure, characteristics of the analysis, outcomes, and ease of adoption. They also enlighten the limitations of adopting the existing approaches and discuss the current state of their adoption in software engineering processes. Their observations indicate that the threat analysis procedures are not clearly defined and they have a lack of quality assurance and tool support. Xiong et al. \cite{xiong2019threat} performed a systematic literature review on threat modeling. They reviewed 54 articles and identified three types of articles among these: (i) articles which are making a contribution (such as introducing a new method) to the field of threat modeling, (ii) articles which are using an existing threat modeling method, and (iii) articles which are presenting work related to the threat modeling process. They observed that most threat modeling work are done manually with a limited assurance of their validation. 

\textbf{Cybersecurity information extraction:} Bridges et al. \cite{bridges2017cybersecurity} evaluated existing methods (\cite{bridges2013automatic, joshi2013extracting, jones2015towards}) of accurate and automatic extraction of security entities from text. They used online news and blog articles, websites of common vulnerabilities and exposures (CVE), National Vulnerability Database (NVD), and Microsoft security bulletins. After comparing the existing approaches, the authors drew conclusions that the existing methods have a low recall, and no large publicly available annotated data set of security documents is available. Overall, these aforementioned researches focus on CTI from various perspectives such as privacy, sharing, modelling, and performance. 

\textbf{Previous work by the authors:} In our previous work~\cite{rahman2020literature}, we conducted a systematic literature review on 34 CTI extraction studies. We identified 8 data sources for collecting CTI candidate texts and 7 CTI extraction purposes. We also identified the natural language processing techniques used for CTI extraction. However, in this study, we expand our previous work by investigating a greater number of relevant studies, identify three new CTI extraction purposes, and propose a CTI extraction pipeline which can be instantiated by prospective researchers for CTI extraction.


\section{Methodology}
\label{methodlogy}
In this section, we explain the process of searching, selecting, and analysing the studies.

\subsection{Search strategy}
\label{search}
The first step is to find relevant studies from the scholar databases. We select six scholar databases to conduct our search: Institute of Electrical and Electronics Engineers (IEEE) Xplore \cite{ieeex}, Association for Computing Machinery (ACM) Digital Library \cite{acmdl}, ScienceDirect \cite{sciencedirect}, SpringerLink \cite{springer}, Wiley Online Library \cite{wiley}, and DBLP \cite{dblp}. We construct a set of search strings to identify relevant studies in the selected scholar databases. We search each of the six scholar databases using the search queries given below. In total, we find 20,922 publications after removing duplicates. 

\begin{enumerate}[label=\alph*), leftmargin=4mm]
    \item (threat \textit{OR} cyber*) \textit{ONEAR/2} (intelligence \textit{OR} action* \textit{OR} advisories)
    \item (threat \textit{OR} cyber* \textit{OR} security) \textit{ONEAR/2} (report* \textit{OR} article* \textit{OR} information \textit{OR} threat*)
    \item "hacker forum*" \textit{OR} "dark*" \textit{OR} "cti" \textit{OR} "tactics, techniques and procedures" \textit{OR} "apt attack*"
\end{enumerate}

\subsection{Selection of relevant studies}
Our search result includes publications that are not relevant to extracting CTI from text. Hence, we establish inclusion and exclusion criteria to filter the irrelevant publications:

\textbf{Exclusion criteria:}
\label{criteria}
\begin{enumerate*}[label=\alph*)]
    \item Publications that are not peer-reviewed: keynote abstracts, call for papers, and presentations; or
    \item publications that were published before 2000; or
    \item publications that are written in languages except in English
\end{enumerate*}

\textbf{Inclusion criteria:}
\begin{enumerate*}[label=\alph*)]
    \item Publications must be available for downloading or reading on the web; and
    \item title, keywords, and abstract of the study, which explicitly indicates that the publication is related to extracting CTI from textual documents.
\end{enumerate*}

The filtering is done by the first two authors of this article. The first author filtered the search results manually through the inclusion and exclusion criteria. The second author used FAST$^2$ \cite{Yu2019}, an intelligent tool for publication selection in literature reviews. Carver et al. \cite{carver2013identifying} show that selecting publications in systematic literature reviews is one of the hardest and most time-consuming tasks. Therefore, using an intelligent tool can make the process of selecting studies more efficient. 
Using FAST$^2$ \cite{Yu2019}, the second author starts selecting key publications from the list of papers found from the search results. These key publications are the relevant publications the authors first studied before searching the scholar database for the relevant studies (Section~\ref{search}). Then a model is trained by the tool based on the title and abstract of the studies. The model ranks the rest of the studies based on their relevance, and a list of ten candidate studies are shown. The second author selects the relevant studies from the list of candidates, and this feedback makes the model more accurate in each iteration. Thus, the tool can predict the number of relevant publications through supervised learning in the list based on the user feedback. Using this predicted number, the tool stops when 95\% of all relevant publications are selected. 

After the first two authors finish the filtering process, we obtain a set of 50 publications. We refer to these 50 studies as the initial set. After a detailed read of these publications, we find studies that are irrelevant. For example, Iqbal et al. \cite{iqbal2018stixgen} suggested an approach that threat analysts \textit{manually} find CTI from text reports and add the information to a database for STIX \cite{stix} generation. With discussions between the first two authors on the relevance of each publication, we select 33 publications from the initial set. We refer to these 33 studies as Study Set A. Then, we apply forward and backward snowballing~\cite{wohlin2012experimentation} on Study Set A. For forward snowballing, we collect publications that cite Study Set A. For backward snowballing we collect the publications that are cited by Study Set A. Afterwards, we apply the exclusion and inclusion criteria (Section ~\ref{criteria}) on these snowballed publications which gets us a new set of 31 publications which we refer to as Study Set B. Finally, combining the Study Set A and B, we get a total of 64 studies which we select for our survey study. We report these  studies at Table~\ref{tab:selected-studies} in the Section~\ref{appendix}. These studies are listed in alphabetical order and referred to as $S_x$, for example, $S_1$ refers to the study \textit{A machine learning-based FinTech cyberthreat attribution framework using high-level indicators of compromise}.

\subsection{Analysing the studies}
\label{coding}
After selecting the studies, the first two authors each read the 64 papers. While reading each publication, they use the open coding technique \cite{saldana2015coding} and take note of the following information from each study: \begin{enumerate*}[label=\alph*)]
    \item the stated CTI extraction purpose,
    \item the details of the data sources, and
    \item the steps followed by the authors in their methodology.
\end{enumerate*} After reading the publications and applying open coding, the first and second authors apply card sorting \cite{spencer2004card} on the three sets of codes mentioned above. Then, they resolve the disagreements and agreed on the following: \begin{enumerate*}[label=\alph*)]
    \item ten types of CTI extraction purposes (see Section~\ref{goal});
    \item seven types of data sources to extract CTI (see Section~\ref{data-source});
    \item the pipeline consisting of six  steps for CTI extraction process as shown in Figure~\ref{fig:pipeline}; and
    \item the categorization of techniques and CTI sharing formats (see Section~\ref{mining-approach}).
\end{enumerate*} We report our findings in the following sections.

\begin{small}
\begin{table}[htb]
    \centering
    \caption{CTI Extraction purposes}
    \begin{tabular}{p{4cm}p{6cm}l}
    \toprule
        \textbf{Purpose (abbreviation*)} & \textbf{Studies} & \textbf{CTI subdomain}  \\
        
        CTI-related text classification (TC) & $S_3$, $S_{12}$, $S_{14}$, $S_{20}$, $S_{21}$,  $S_{26}$, $S_{30}$, $S_{31}$, $S_{32}$, $S_{38}$, $S_{39}$, $S_{52}$ & Strategic\\ 
        
        Attack tactics, techniques, and procedures extraction (TTP) & $S_1$, $S_2$, $S_5$, $S_8$, $S_{15}$, $S_{35}$, $S_{37}$, $S_{55}$, $S_{60}$, $S_{62}$ & Tactical \\ 
        
        Cybersecurity-related keywords extraction (KW) & $S_9$, $S_{18}$, $S_{23}$, $S_{29}$, $S_{40}$, $S_{44}$, $S_{51}$, $S_{53}$, $S_{61}$ & Strategic\\ 
        
        Cybersecurity-related event identification (EI) & $S_4$, $S_{16}$, $S_{17}$, $S_{22}$, $S_{36}$, $S_{45}$, $S_{54}$, $S_{56}$, $S_{64}$ & Strategic \\ 
        
        Indicators of compromise extraction (IoC) & $S_6$, $S_{11}$, $S_{13}$, $S_{42}$, $S_{46}$, $S_{57}$, $S_{58}$, $S_{59}$ & Technical \\ 
        
        Cyberthreat alert generation (AG) & $S_{19}$, $S_{24}$, $S_{25}$, $S_{49}$, $S_{63}$ & Operational\\ 
        
        Hacker resource analysis (HR) & $S_{10}$, $S_{27}$, $S_{28}$,  $S_{41}$, $S_{43}$ & Strategic\\ 
        
        Software vulnerability information extraction (SV) & $S_{33}$, $S_{34}$, $S_{47}$, $S_{48}$ & Technical\\ 
        
        Threat report classification (CL) & $S_7$ & Tactical\\
        
        Cyberthreat actor attribution (AA) & $S_{50}$ & Strategic\\ 
    \bottomrule    
    \multicolumn{3}{c}{*These are short names for each of the CTI extraction purposes} \\
    \end{tabular}
    \label{tab:cti-goals}
\end{table}
\end{small}

\section{CTI extraction purpose}
\label{goal}
Using the methodology described in Section~\ref{coding}, we identified ten purposes for extracting CTI information. The agreement score (Cohen's Kappa~\cite{mchugh2012interrater}) is $0.34$. We discuss these purposes in the following subsections and summarize the purposes with citations to the studies in Table~\ref{tab:cti-goals}. 

\subsection{CTI-related text classification (TC) (n=12)}
To extract information from the CTI-candidate texts (textual sources for collecting CTI information), researchers first need to determine whether these texts are CTI related. For example, to extract IoCs from a list of Twitter posts, the CTI-related posts need to be filtered from the non-relevant posts through text classification. For example, in $S_{12}$, $S_{20}$, $S_{30}$, $S_{31}$, $S_{32}$, $S_{39}$, and $S_{52}$, the authors classified whether the hacker forum posts are CTI related. In $S_3$, $S_{14}$, $S_{21}$, $S_{26}$, and $S_{38}$ the authors determined whether social media posts and online articles are CTI related. Classifying CTI relevant texts serves as a precursor to other CTI extraction purposes, such as IoC or TTPs extraction. Because the process of extracting IoCs, attack patterns, cyberthreat events information from text includes the process of CTI related text classification and the gained insight can also be used as strategic CTI. 

\subsection{Attack tactics, techniques, and procedures extraction (TTP) (n=10)} Threat reports contain verbose details on how malicious attacks are being performed. Attack tactics, techniques, and procedure (TTP)-related information can be extracted from these textual reports. For example, in $S_1$, $S_2$, $S_5$, $S_8$, $S_{35}$, and $S_{55}$, attack patterns are identified from threat reports. In $S_{37}$, the authors developed an automated feature engineering mechanism to identify malware propagation tactics and techniques from the security literature. In $S_{15}$, the authors extracted the malware actions from the threat reports and represented these actions in a cybersecurity knowledge graph. In $S_{60}$ and $S_{62}$, the authors developed an approach to extract threat actions from threat reports using their threat-action ontology, which refers to a set of categories, concepts, properties, and their relationship with one another in a subject domain~\cite{ontology-wiki}. Then, they mapped each threat action to the appropriate techniques, tactics, and kill chain phases\footnote{\url{https://www.lockheedmartin.com/en-us/capabilities/cyber/cyber-kill-chain.html}}. Extracted TTPs serve as tactical CTI and help the defending organizations plan to prevent themselves from being attacked. 

\subsection{Cybersecurity-related keywords extraction (KW) (n=9)}
Cybersecurity-relevant keywords, such as name of malware, affected systems, applications, vulnerability, exploits, and files, can be extracted from text. For example, in $S_{18}$, $S_{44}$, $S_{51}$, the authors extracted the organization, application, version number, and vulnerability names from the Twitter posts, NVD vulnerability description, and online articles, respectively. In $S_9$, $S_{23}$, $S_{29}$, and $S_{40}$, the authors extracted the trending names of attacks, hacking tools, and exploits from the hacker forum posts such as data breach, denial of service, botnet and SQL injection. In $S_{53}$, the authors extracted the trending keywords from the Twitter posts. In $S_{61}$, the authors extracted the trending keywords from security blog posts. Similar to the CTI-related text classification, extracting cybersecurity related keywords also serves as a precursor to other CTI extraction purposes, such as IoC or TTPs extraction. The process of extracting IoCs, attack patterns, and cyberthreat event information from text includes the process of CTI-related text classification and keyword extraction. Therefore, the insight gained from cybersecurity-related keyword extraction can be used as strategic CTI. 

\subsection{Cybersecurity-related event identification (EI) (n=9)} Text mining can be applied to identify whether CTI-related texts are describing cybersecurity-related events or incidents. In the  studies, Twitter is explored mostly for identifying cyberthreat incident-related posts and extracting the incident-related information in  $S_4$, $S_{16}$, $S_{17}$, $S_{22}$, $S_{36}$, $S_{45}$, $S_{56}$, and $S_{64}$. Moreover, in $S_{54}$, the authors extracted cybersecurity event information from online articles. These extracted threat-related events serve as strategic CTI, which can help cybersecurity practitioners and researchers to understand past events, monitor future threats, and share the CTI information with other cybersecurity information-consuming organizations. 
    
\subsection{Indicators of compromise extraction (IoC) (n=8)} CTI-related texts contain information on indicators and traces of compromise, which provide evidence of malicious activities or compromises in systems. Extracting these indicators of compromise (IoC) such as malware names, signatures, hashes, IP addresses, and packets facilitates further research and analysis opportunities for security research and practitioners. In the  studies, IoCs have been extracted from social media ($S_{46}$, $S_{58}$), threat reports ($S_6$, $S_{57}$, $S_{59}$), and online articles ($S_{13}$, $S_{42}$). Moreover, in $S_{11}$, specific indicators of malware delivery attempts by advanced persistent threat (APT) attack groups have been extracted from threat reports. These extracted IoCs serve as technical CTI and can be utilized to learn about future attack attempts, analyzing malware behavior, and designing anti-malware tools.
    
\subsection{Cyberthreat alert generation (AG) (n=5)} Alerts on emerging attack patterns, tools, malware, and vulnerabilities can be generated from CTI-related texts. For example, in $S_{19}$, and $S_{63}$, the authors have generated threat alerts based on Twitter posts by security experts. Moreover, darknet and hacker community discussion threads have been explored to warn about future cyberattacks in $S_{24}$, $S_{25}$, and $S_{49}$. These warnings can be generated for target organizations and industries which can serve as operational CTI. 
    
\subsection{Hacker resource analysis (HR) (n=5)} Darknet and hacker forums are rich in hacking tools, documents, scripts, source code, and online resources regularly being used by the attackers' community. Researchers have analyzed these resources in their studies to gain further insight on the attackers motives and techniques. For example, In $S_{10}$, $S_{27}$, $S_{28}$, $S_{41}$, and $S_{43}$, Samtani, Williams, and Grisham et al. proposed a platform named AZSecure, which facilitates security researchers and practitioners to search and visualize cutting-edge hacking tools, scripts, malware and other relevant assets to launch cyberattacks. Extracting CTI from these sources serves as strategic CTI and provides insight on cutting-edge hacking tools, trends on emerging threat-related issues.
    
\subsection{Software vulnerability information extraction (SV) (n=4)} Open source repositories of version control systems contain information on vulnerable software packages, insecure development practices, security bugs, issues, and developers' discussions. Software vulnerability-related information can be extracted to gain insight on securing the systems proactively. For example, in $S_{47}$, the authors have extracted vulnerable package-related information from Github repositories. In $S_{48}$, the authors extracted vulnerability trends and patterns from the  CVE from the NVD\cite{nvdcve} description. In $S_{33}$, $S_{34}$, Mulwad et al., and Joshi et al. extracted the vulnerability and related keywords from vulnerability description in NVD, extracted their underlying concepts using ontology and online resource descriptors to facilitate cybersecurity practitioners to query and reason over those linked concepts using graph-based knowledge base. This vulnerability information can serve as technical CTI and can be used to proactively secure software artifacts from vulnerable artifacts. 

\subsection{Threat report classification (CL) (n=1)} Threat reports can be classified based on the types of threats being discussed in the report. For example, in $S_7$, the authors classified the threat reports based on the name of the malware discussed in the report. These classified threat reports can serve as tactical CTI.

\subsection{Cyberthreat actor attribution (AA) (n=1)} CTI candidate texts often contain information on cyberthreat incidents and associated cyberattack actors, such as their roles, strategies, and procedures. These information can be used to map the attacks to the responsible cyberthreat actors. For example, in $S_{50}$, the authors identified the responsible attack groups from the threat reports, where the group(s) are responsible for launching cyberattacks through propagating malware. Thus, cyberthreat actor attribution serves as a strategic CTI and can be utilized to gain tactical insight into the attackers' strategies and malicious activities.

\begin{figure*}[h]
\centering
\includegraphics[width=0.8\columnwidth]{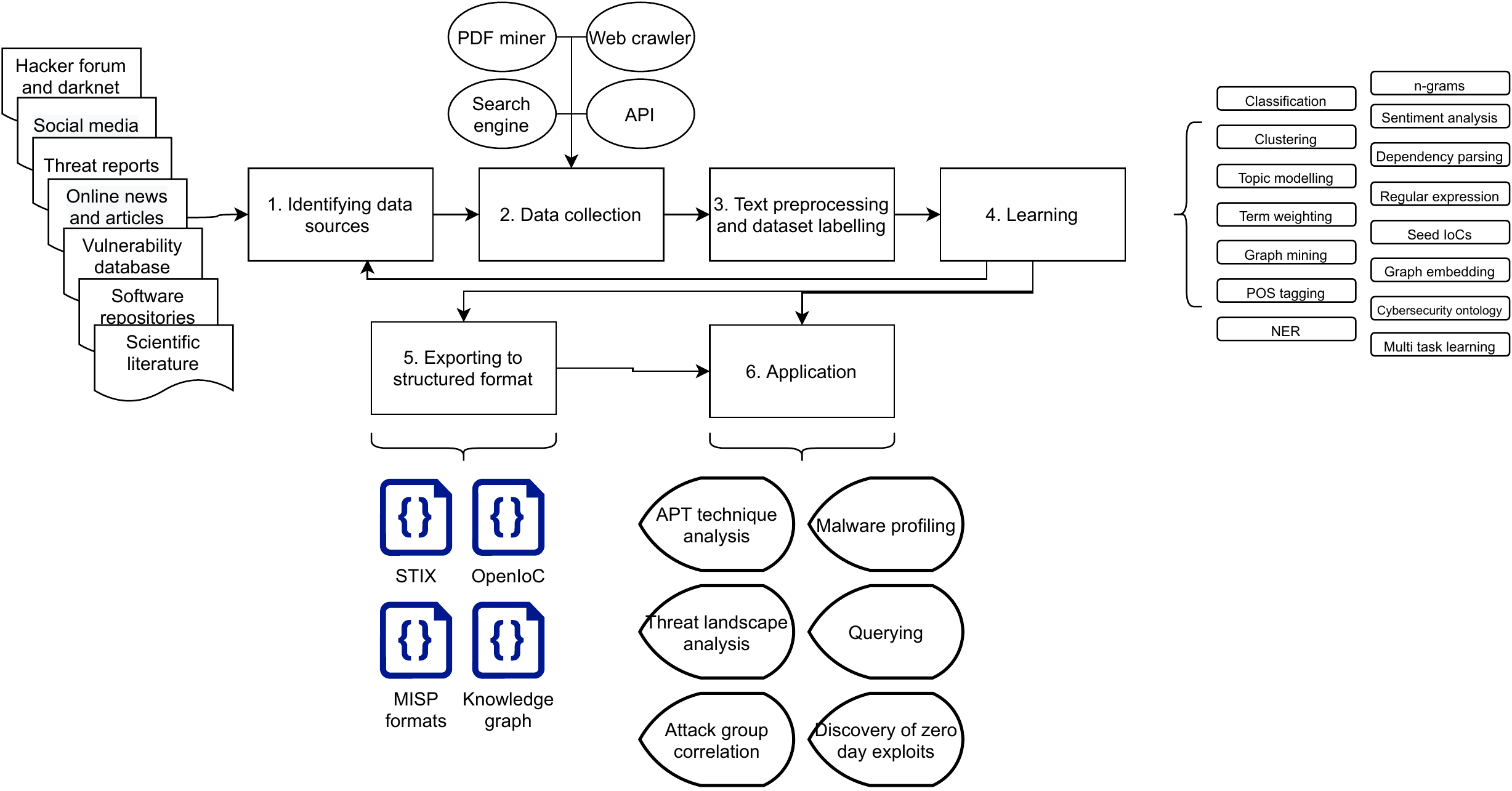}
\caption{CTI extraction pipeline}
\label{fig:pipeline}
\end{figure*}

\section{CTI extraction pipeline}
\label{pipeline}
Extracting CTI-related information from CTI-candidate text requires NLP in combination with ML techniques. Although the  extraction techniques depend on the types of information being extracted from the text (e.g., the extraction procedure of IoCs can be different than that of TTPs), the abstracted pipeline stays similar. Hence, we propose a CTI extraction pipeline, which abstracts the extraction approaches found in the studies. Thus, the pipeline can be instantiated based on the CTI extraction purposes and provide cybersecurity researchers with possible options to design their own CTI extraction pipeline. Using the methodology described in Section~\ref{coding}, we identify six steps, and the agreement score is $1.00$. In Figure~\ref{fig:pipeline}, we show the steps of the pipeline. The pipeline has these six following steps: (i) data source identification, (ii) data collection, (iii) text preprocessing and dataset labelling, (iv) learning, (v) exporting to structured formats, and (vi) application. The fifth and sixth steps do not exist in all studies, such as in $S_9$ and $S_{13}$, however, the steps regarding data collection, labelling, and learning are common in all studies. In the following subsections, we describe each of these steps. 


\begin{small}

\begin{longtable}{llp{5cm}p{4cm}l}
    \caption{Data sources for CTI extraction in the studies} 
    \label{tab:data-source} \\ \hline
        \textbf{Study} & \textbf{Purpose} & \textbf{Sources} & \textbf{Type} & \textbf{Instances}  \\
        
        $S_1$ & TTP & APT Notes \cite{aptnotes, cybercriminal} & Threat report & 327 \\
        
        $S_2$ & TTP & APT Notes & Threat report & 445 \\
                
        $S_3$ & TC & Twitter\cite{twitter} & Social media & 200K \\
                
        $S_4$ & EI & Twitter\cite{cybertweets} & Social media & 21K \\
        
        $S_5$ & TTP & FireEye\cite{fireeye}, Kaspersky Security Lab \cite{kaspersky}, Apt Notes  & Threat report & 50 \\
        
        $S_6$ & IOC & AlienVault (AT\&T CyberSecurity Blog) \cite{alienvault}, FireEye, MalwareBytes \cite{malwarebytes} & Online blogs and articles & 71K \\
        
        $S_7$ & CL & Threat Expert* & Threat report & 25K \\
        
        $S_8$ & TTP & Symantec\cite{symantec}, FireEye, MacAfee \cite{mcafee} & Threat report & 18K \\
        
        $S_9$ & KW & FireEye, Krebs On Security \cite{krebsonsecurity}, Securelist\cite{kaspersky}, F-secure\cite{fsecure}, Crowdstrike\cite{crowdstrike} & Online blogs and articles & 13K \\
        
        $S_{10}$ & HR & OpenSC* & Hacker and darknet forum & 5K \\
        
        $S_{11}$ & IOC & WeLiveSecurity \cite{welivesecurity}, TaoSecurity \cite{taosecurity}, Malwarebytes\cite{malwarebytes} , Trend Micro \cite{trendmicro} & Online blogs and articles & 14K \\
        
        $S_{12}$ & TC & nulled.io* & Hacker and darknet forum & 16K \\
        
        $S_{13}$ & IOC & APT Notes  & Threat reports & 687 \\
        
        $S_{14}$ & TC & Twitter & Social media & 21K \\
        
        $S_{15}$ & TTP& FireEye, Kaspersky  & Threat report & 474 \\
        
        $S_{16}$ & EI & Twitter & Social media & 5.1B \\
        
        $S_{17}$ & EI & Recorded Future.com~\cite{recordedfuture}, FireEye, Security Week~\cite{securityweek}, Trend Micro& Threat report & \textit{NM}\footnote{NM stands for "not mentioned" in the corresponding study} \\
        
        $S_{18}$ & KW & Twitter~\cite{relevanttweets} & Social media & 11K \\
        
        $S_{19}$ & AG & Twitter & Social media & 144K \\
        
        $S_{20}$ & TC & \textit{NM} & Hacker and darknet forum & 5.4k \\
        
        $S_{21}$ & TC & Twitter & Social media & 195K \\
        
        $S_{22}$ & EI & Twitter~\cite{cyberthreatdetectiondata} & Social media & 5K \\
        
        $S_{23}$ & KW & CrimeBB* & Hacker and darknet forum & 42M\\
        
        $S_{24}$ & AG & Twitter, web crawler\footnote{the CTI candidate texts are collected by a web crawler, however the name/source of the crawler have not been mentioned} & Social media, online blogs and articles, hacker and darknet forum & 764 \\
        
        $S_{25}$ & AG & Twitter, \textit{NM} & Social media, hacker and darknet forum & - \\
        
        $S_{26}$ & TC & Twitter, Cracking Arena*, Dream Market* & Social media, hacker and darknet forum & 148K \\
        
        $S_{27}$ & HR & OpenSC* & Hacker and darknet forum & 432K \\
        
        $S_{28}$ & HR & Hackfive*, Hackhound*, Icode* & Hacker and darknet forum & 672K \\
        
        $S_{29}$ & KW & Cracking Fire* & Hacker and darknet forum & 38K \\
        
        $S_{30}$ & TC & Sixgill Crawler \cite{sixgill} & Hacker and darknet forum & 3K \\
        
        $S_{31}$ & TC & \textit{NM} & Hacker and darknet forum & 1.3M\\
        
        $S_{32}$ & TC & nulled.io* & Hacker and darknet forum & 16K \\
        
        $S_{33}$ & SV & Microsoft security bulletins~\cite{microsoftsecurity}, Adobe security bulletins~\cite{adobesecurity}, CVE Description~\cite{nvdcve} & Threat reports, vulnerability database, online blogs and articles & 320 \\
                
        $S_{34}$ & SV & CVE Description, CNET \cite{cnet} & Vulnerability database, online blogs and articles & 155 \\
                
        $S_{35}$ & TTP & \textit{NM} & Threat reports & \textit{NM} \\
        
        $S_{36}$ & EI & Twitter & Social media & 5B \\
        
        $S_{37}$ & TTP & IEEE Symposium of Security and Privacy (2008-15)~\cite{IEEESP}, IEEE Computer Security Foundation Symposium (2000-14)~\cite{IEEECSF}, USENIX Security Symposium (2011)~\cite{usenix} & Scientific literature & 1K \\
        
        $S_{38}$ & TC & Twitter, CVE Description & Social media, vulnerability database & 76K \\
        
        $S_{39}$ & TC & Cracking Arena* & Hacker and darknet forum & 45K \\
        
        $S_{40}$ & KW & \textit{NM} & Hacker and darknet forum & 26K \\
                
        $S_{41}$ & HR & Hackhound*, Ashiyane*, VBSpiders*, Zloy* & Hacker and darknet forum & 482K \\
        
        $S_{42}$ & IOC & FireEye, Kaspersky & Threat report & 400 \\
        
        $S_{43}$ & HR & OpenSc*, Cracking Zilla*, AntiOnline \cite{antionline} & Hacker and darknet forum & 1.3M \\
        
        $S_{44}$ & KW & CVE Description & Vulnerability database & 20k \\
                
        $S_{45}$ & EI & The Hacker News~\cite{thehackernews} & Online blogs and articles & - \\
        
        $S_{46}$ & IOC & Twitter & Social media & 2.3k \\
        
        $S_{47}$ & SV & Github \cite{github} & Software repository & 111K \\
        
        $S_{48}$ & SV & CVE Description & Vulnerability database & 26K \\
        
        $S_{49}$ & AG & Dream Market*, Darkshades market* & Hacker and darknet forum & 48K \\
        
        $S_{50}$ & AA & Attack Attribution Dataset \cite{attackattribution} & Threat report & 249 \\
        
        $S_{51}$ & KW & \textit{NM} & Online blogs and articles & 30K\\
                
        $S_{52}$ & TC & \textit{NM} & Hacker and darknet forum & 33K\\
                
        $S_{53}$ & KW & Twitter & Social media & 539K \\
                
        $S_{54}$ & EI & Freebuf~\cite{freebuf}, Easyaq~\cite{easyaq} & Online blogs and articles & 610 \\
        
        $S_{55}$ & TTP & Microsoft Security Advisory~\cite{microsoftsecurityadvisory}, Cisco Security Advisories~\cite{ciscosecurityadvisory}, Crowdstrike & Threat report & \textit{NM} \\
        
        $S_{56}$ & EI & Twitter & Social media & 47.8M \\
        
        $S_{57}$ & IOC & \textit{NM} & Threat report & 190 \\
        
        $S_{58}$ & IOC & AlienVault, FireEye, Microsoft security bulletins, Cisco security bulletins, Kaspersky, Webroot~\cite{webroot}, Hack Forums\cite{hackforums} & Social media, online blogs and articles, hacker and darknet forum & 15k \\
        
        $S_{59}$ IOC & & Threat Expert* & Threat report & 25K \\
        
        $S_{60}$ & TTP & Symantec & Threat reports & 17K \\
        
        $S_{61}$ & KW & Cisco \cite{cisco}, Symantec, FireEye, Palo Alto \cite{paloaltonetworks} & Threat report & 875 \\
        
        $S_{62}$ & TTP & Symantec & Threat report & 2.2K \\
        
        $S_{63}$ & AG & Twitter & Social media & 1.1B \\
                
        $S_{64}$ & EI & Twitter & Social media & 15M \\  
        
        \hline
        
        \multicolumn{4}{c}{Sources marked with an asterisk (*) indicate that the url are not found} \\
    
\end{longtable}
\end{small}

\subsection{Step 1: identifying data sources}
\label{data-source}
The process for extracting CTI starts with identifying data sources for CTI candidate texts. Using the methodology described in Section~\ref{coding}, we identify seven data source categories, and the agreement score is $1.00$. The data source categories are described below and are reported in Table~\ref{tab:data-source}. 

\subsubsection{Hacker and darknet forums (n=20)}
Online forums maintained by security vendors, experts, ethical hackers, penetration testers, and malicious users provide a platform for cyberthreat-related knowledge sharing. These forums, such as Cracking Arena, OpenSC, and Crackingzilla, facilitate hacker communities to share attack tactics, techniques, malicious scripts, and tools, and security best practices. Similarly, hacker forums can be hosted in the darkweb (such as Dream Market) which provides the aforementioned CTI-related information. Hence, CTI information can be extracted from these forum discussion and attachments, such as CTI related text classification ($S_{12}$, $S_{20}$, $S_{26}$, $S_{30}$, $S_{31}$, $S_{32}$, $S_{39}$, $S_{52}$), hacker resource analysis ($S_{10}$, $S_{27}$, $S_{28}$, $S_{42}$, $S_{43}$), cyberthreat alert generation ($S_{24}$, $S_{25}$, $S_{49}$), CTI-related keywords extraction ($S_{23}$, $S_{29}$, $S_{40}$), and  indicators of compromise extraction ($S_{58}$).

\subsubsection{Social media (n=19)}
In recent years, social media has become an important source for news and updates related to literally every aspect of life and CTI is no exception. Cybersecurity researchers and vendors post cutting edge news and updates on threats, attacks, and zero-day exploits. Reputed security experts and organizations, i.e., Brian Krebs \cite{krebsonsecurity} and InterSec WorldWide \cite{intersecww}, regularly post feeds related to CTI and security incidents on Twitter. Social media posts have been used for CTI extraction, such as cybersecurity related event identification ($S_4$, $S_{16}$, $S_{22}$, $S_{36}$, $S_{56}$, $S_{64}$); CTI related text classification ( $S_3$, $S_{14}$, $S_{21}$, $S_{26}$, $S_{38}$), cyberthreat alert generation ($S_{19}$, $S_{24}$, $S_{25}$; $S_{63}$); cybersecurity-related keywords extraction ($S_{18}$, $S_{53}$); and  IoC ($S_{46}$, $S_{58}$).

\subsubsection{Threat reports (n=18)}
Threat reports are usually prepared by reputed cybersecurity vendors who regularly publish threat reports on the current threat landscape, advanced persistent attacks, corresponding attack groups along with their working strategies and TTPs. These reports also cover vulnerabilities and exploits for specific technologies, cyberthreat events such as data breaches and zero-day attacks. Reputed cybersecurity vendors such as Fireeye \cite{fireeye}, Symantec \cite{symantec}, Kaspersky Lab \cite{kaspersky}, Trend Micro \cite{trendmicro-tr} are regularly providing aforementioned analysis through textual reports. Threat reports have been used for CTI extraction, such as attack tactics, techniques, and procedures extraction ($S_1$, $S_2$, $S_5$, $S_8$, $S_{15}$,  $S_{35}$,  $S_{55}$, $S_{60}$, $S_{62}$ ), indicators of Compromise extraction ($S_{13}$, $S_{42}$, $S_{57}$, $S_{59}$), cyberthreat actor attribution ($S_{50}$), cybersecurity related event identification ($S_{17}$), cybersecurity related keywords extraction ($S_{61}$), software vulnerability information extraction ($S_{33}$), and threat report classification ($S_7$).

\subsubsection{Online blogs and articles (n=10)}
With the increase of cyberattacks, the number of cybersecurity-related web articles and blogs are increasing, which discuss cutting-edge attack techniques, malware descriptions, and attack prevention guidelines. They also provide information on threat prevalence and distribution. These blogs and web articles are written by individuals, communities, or organizations having expertise in cybersecurity domains. We Live Security, Krebs on Security, Threatpost, AlienVault are examples of reputed cybersecurity related blogs and websites. Threat reports have been used for CTI extraction, such as indicators of compromise extraction ($S_6$, $S_{11}$, $S_{58}$), cybersecurity-related event identification ($S_{45}$, $S_{54}$),  cybersecurity-related keywords extraction ($S_9$, $S_{51}$), software vulnerability information extraction ($S_{33}$, $S_{34}$), and cyberthreat alert generation ($S_{24}$).

\subsubsection{Vulnerability databases (n=5)}
The NVD \cite{nvd} is the U.S. government's repository of known vulnerabilities in software and hardware components that are described as common vulnerabilities and exposures (CVEs) \cite{nvdcve}. The type of each vulnerability is determined by assigning a common weakness enumeration (CWE)~\footnote{\url{https://cwe.mitre.org/index.html}} number.  The CWE is a list of software and hardware weakness types developed by the security community. Researchers extract CTI from the vulnerability descriptions in CVEs. Moreover, several researchers, such as Husari et al. ($S_{60}$), use the Att\&CK Framework \cite{attack} and Common Attack Pattern Enumeration and Classification (CAPEC) \cite{capec} to develop a threat ontology. The ATT\&CK Framework is a globally-accessible knowledge base of real-world attacks that contains TTPs including tactics.  Each tactic has a list of techniques, and each technique has procedure examples. CAPEC is a comprehensive dictionary of known attack patterns used by adversaries to exploit known vulnerabilities. Vulnerability databases have been utilized to extract CTI such as software vulnerability information extraction ($S_{33}$, $S_{34}$, $S_{48}$), cybersecurity-related keywords extraction ($S_{44}$), CTI-related text classification ($S_{38}$).

\subsubsection{Software repositories (n=1)}
Software engineering practitioners use version control repositories to store software development artifacts, such as source code, bugs and issues, wiki, and documentation. Github \cite{github}, GitLab \cite{gitlab}, and BitBucket \cite{bitbucket} are examples of popular version control repositories. These repositories contain textual information on security bugs, vulnerable packages, and developers' discussions on pertinent security topics. Software repositories have been utilized to extract CTI such as software vulnerability information extraction ($S_{47}$).

\subsubsection{Scientific literature (n=1)}
Scientific literature published in cybersecurity conferences and journals contains scientific observation and analysis on cyberthreat-related issues, such as malware behaviors, cyberattack approaches, and attack mitigation techniques. These textual documents can thus be the source for extracting CTI relevant information. Zhu et al. ($S_{37}$) extracted Android malware indicators and behaviors from reputed cybersecurity venues named IEEE Security and Privacy, USENIX Security, and Computer Security Foundation Symposium. 

\subsection{Step 2: data collection}
\label{data-collection}
From the sources mentioned in Table~\ref{tab:data-source}, data can be collected in a manual or automated manner. Researchers can manually collect CTI-candidate texts. However, this method is error-prone and inefficient. Hence, researchers used online search engines to find relevant articles, threat reports ($S_{42}$). Additionally, custom search engines for finding cybersecurity articles have also been explored in $S_1$, $S_{30}$. However, using search engines to find relevant articles can be time consuming, and hence, researchers extensively used web-based crawlers to find online cybersecurity related topics, especially in hacker forums and darknet web resources. In $S_6$, $S_{11}$, $S_{13}$, $S_{20}$, $S_{31}$, $S_{43}$, $S_{51}$, $S_{54}$, researchers used HTML-based web crawlers to identify relevant texts. Moreover, TOR browser\footnote{https://www.torproject.org/} based crawlers are used in $S_{10}$, $S_{24}$, $S_{25}$, $S_{27}$, $S_{49}$, $S_{52}$ to find relevant topics in the darkweb. Forum web crawlers, such as Sixgill and Open Discussion Forum Crawler (ODFC)~\cite{kadoguchi2019exploring}, are used in $S_{30}$, $S_{40}$. Text from the threat reports in pdf format can be extracted using tools such as PDFMiner \cite{pdfminer}. This technique is used to scrap the text from pdf documents in $S_{15}$, $S_{37}$, $S_{60}$. Finally, social networking websites and software repositories provide application programming interfaces (API) to collect data from their websites as well. The Twitter API\footnote{https://developer.twitter.com/en/docs/twitter-api} and Github API\footnote{https://docs.github.com/en/rest} are used for collecting tweets and software development artifacts in $S_3$, $S_4$, $S_{14}$, $S_{18}$, $S_{19}$, $S_{21}$, $S_{22}$, $S_{47}$, $S_{53}$, $S_{56}$, $S_{63}$. In $S_1$, $S_2$, $S_5$, $S_{13}$, threat reports available on one of the two GitHub repositories \cite{aptnotes, cybercriminal} are also used. 

\subsection{Step 3: text preprocessing and dataset labelling}
\begin{itemize}[leftmargin=3mm]
    \item \textbf{Substep-3.1 Preprocessing:} Data collected from the mentioned sources in Section~\ref{data-source} are in textual format. Before extraction of CTI, these texts need to be preprocessed to discard punctuation marks, URLs, irrelevant symbols, stop words, and incorrect spellings. Then, tokenization and lemmatization are also applied. These preprocessings are needed to prepare the textual data suitable for NLP techniques.
    \item \textbf{Substep-3.2 Labeling:} The data needs to be labelled for building training and test cases for the ML models. For example, in the case of CTI-relevant text classification, text segments are labelled whether they are CTI relevant or not. In the case of IoC extraction, words in the text are annotated as the type of the indicators. Correct labelling of the dataset is important as the performance of the ML models depends on the correctness of the labelling done by humans. In the studies, researchers labelled the data themselves or deployed multiple annotators who have expertise in cybersecurity, such as graduate students. In $S_1$, $S_4$, $S_5$, $S_7$, $S_8$, $S_9$,  $S_{11}$,  $S_{14}$, $S_{21}$, $S_{26}$, $S_{32}$, $S_{33}$, $S_{38}$, $S_{39}$, $S_{50}$, $S_{58}$, $S_{60}$, the researchers utilized manual labelling. However, the usage of prelabelled datasets can omit the need for manual annotation such as in $S_{16}$, $S_{36}$ where they used the prelabelled dataset as the ground truth named Hackmageddon\cite{hackmageddon} and PrivacyRights\cite{privacyrights}. 
\end{itemize}

\begin{small}

\begin{longtable}{l c |c *{14}{|c}}
    \caption{Extraction techniques} 
    \label{tab:extraction_techniques} \\ \hline
         &
        \textbf{Studies} & 
        \rotatebox[origin=l]{90}{Classification} & 
        \rotatebox[origin=l]{90}{Clustering} & 
        \rotatebox[origin=l]{90}{Topic modelling} & 
        \rotatebox[origin=l]{90}{Term weighting} & 
        \rotatebox[origin=l]{90}{Graph embedding} & 
        \rotatebox[origin=l]{90}{POS-tagging} & 
        \rotatebox[origin=l]{90}{NER} & 
        \rotatebox[origin=l]{90}{Sentiment analysis} & 
        \rotatebox[origin=l]{90}{Regular expression} & 
        \rotatebox[origin=l]{90}{Dependency parsing} & 
        \rotatebox[origin=l]{90}{Seed IoC} & 
        \rotatebox[origin=l]{90}{Graph mining} & 
        \rotatebox[origin=l]{90}{Ontology} & 
        \rotatebox[origin=l]{90}{Multi task learning} & 
        \rotatebox[origin=l]{90}{Historical n-gram} \\
        \midrule 
        \multirow{12}*{\rotatebox[origin=c]{90}{TC}} & $S_3$ & \checkmark & & & & & & & & & & & & & &  \\
        \cline{2-17}
        & $S_{12}$ & \checkmark & & \checkmark & & & & & & & & & & & &   \\
        \cline{2-17}
        & $S_{14}$ & \checkmark & & \checkmark & & & & & & & & & & & &  \\
        \cline{2-17}
        & $S_{20}$ & \checkmark & & & & & & & & & & & & & &   \\
        \cline{2-17}
        & $S_{21}$ & \checkmark & \checkmark & & & & & & & & & & & & &   \\
        \cline{2-17}
        & $S_{26}$ & \checkmark & & & & & & & & & & & & & &  \\
        \cline{2-17}
        & $S_{30}$ & \checkmark & & & & & & & & & & & & & &   \\
        \cline{2-17}
        & $S_{31}$ & & & & \checkmark & & & & & & & & & & &    \\
        \cline{2-17}
        & $S_{32}$ & \checkmark & & & & & & & & & & & & & &    \\
        \cline{2-17}
        & $S_{38}$ & \checkmark & & & & & & & & & & & & & &   \\
        \cline{2-17}
        & $S_{39}$ & \checkmark & & & & & & & & & & & & & &   \\
        \cline{2-17}
        & $S_{52}$ & & & & & \checkmark & & & & & & & & & &   \\
        \midrule

        \multirow{10}*{\rotatebox[origin=c]{90}{TTP}} & $S_1$ & & & \checkmark & & & & & & & & & & & & \\
        \cline{2-17}
        & $S_2$ & & & & & & & & & & \checkmark & & & & &    \\ 
        \cline{2-17}
        & $S_5$ & & & & & & & \checkmark & & & & & & & &    \\
        \cline{2-17}
        & $S_8$ & \checkmark & & & & & & & & & & & & & &  \\
        \cline{2-17}
        & $S_{15}$ & & & & & & & \checkmark & & & & & & \checkmark & &  \\
        \cline{2-17}
        & $S_{35}$ & & & \checkmark & & & & & & & & & & & &   \\
        \cline{2-17}
        & $S_{37}$ & & & & & & & \checkmark & & & \checkmark & & & & &    \\
        \cline{2-17}
        & $S_{55}$ & & & & & & & \checkmark & & & \checkmark  & & & & &   \\
        \cline{2-17}
        & $S_{60}$ & & & & & & & & & & \checkmark & & & \checkmark & &   \\
        \cline{2-17}
        & $S_{62}$ & & & & & & & & & & \checkmark & & & & &   \\
        \midrule
        
        \multirow{9}*{\rotatebox[origin=c]{90}{KW}} & $S_9$ & & & & & & \checkmark & & & & & & & & &  \\
        \cline{2-17}
        & $S_{18}$ & & & & & & & \checkmark & & & & & & & &   \\
        \cline{2-17}
        & $S_{23}$ & & & & \checkmark & & & & & & & & & & &   \\
        \cline{2-17}
        & $S_{29}$ & & & \checkmark & & & & & & & & & & & &   \\
        \cline{2-17}
        & $S_{40}$ & & & & & & \checkmark & & \checkmark & & & & & & &   \\
        \cline{2-17}
        & $S_{44}$ & & & & & & & \checkmark & & & & & & & &   \\
        \cline{2-17}
        & $S_{51}$ & & & & & & \checkmark & \checkmark & & & & & & & &   \\
        \cline{2-17}
        & $S_{53}$ & & & & \checkmark & & & & & & & & & & &   \\
        \cline{2-17}
        & $S_{61}$ & & & \checkmark & & & & & & & & & & & &   \\
        \midrule
        
        \multirow{9}*{\rotatebox[origin=c]{90}{EI}} &  $S_4$ & & \checkmark & & & & & \checkmark & & & & & & & &   \\
        \cline{2-17}
        & $S_{16}$ & & \checkmark & & & & & & & & \checkmark & & & & &  \\
        \cline{2-17}
        & $S_{17}$ & \checkmark & & \checkmark & & & & \checkmark & & & & & & & &  \\
        \cline{2-17}
        & $S_{22}$ & \checkmark & & \checkmark & & & & \checkmark & & & & & & & &   \\
        \cline{2-17}
        & $S_{36}$ & & & & & & & & & & & & & & \checkmark &  \\
        \cline{2-17}
        &  $S_{45}$ & \checkmark & & & & & & & & & & & & & &   \\
        \cline{2-17}
        & $S_{54}$ & \checkmark & & & & & & \checkmark & & & & & & & &  \\
        \cline{2-17}
        & $S_{56}$ & & \checkmark & & & & & & & & & & & & &   \\
        \cline{2-17}
        & $S_{64}$ & \checkmark & & & & & & & & & & & & & &   \\
        \midrule
        
        \multirow{8}*{\rotatebox[origin=c]{90}{IoC}} & $S_6$ & & & & & & & \checkmark & & \checkmark & \checkmark & & \checkmark & & &   \\
        \cline{2-17}
        & $S_{11}$ & & & & & & & \checkmark & & \checkmark & \checkmark & & & & &   \\
        \cline{2-17}
        & $S_{13}$ & & & & & & & \checkmark & & \checkmark & & & & & &   \\
        \cline{2-17}
        & $S_{42}$ & \checkmark & & & & & & & & \checkmark & & \checkmark & & & &  \\
        \cline{2-17}
        & $S_{46}$ & & & & & & & & & \checkmark & & & & & &   \\
        \cline{2-17}
        &  $S_{57}$ & & & & & & & \checkmark & & & \checkmark  & & & & &   \\
        \cline{2-17}
        & $S_{58}$ & & & & & & & \checkmark & & \checkmark & \checkmark & & & & &  \\
        \cline{2-17}
        & $S_{59}$ & & \checkmark & & & & & & & & & & & & &   \\
        \midrule
        
        \multirow{5}*{\rotatebox[origin=c]{90}{AG}} & $S_{19}$ & & & & & & & \checkmark & & & & & & \checkmark & &   \\
        \cline{2-17}
        & $S_{24}$ & & & & \checkmark & & & & & & & & & & &   \\
        \cline{2-17}
        & $S_{25}$ & & & & \checkmark & & & & & & & & & & &   \\
        \cline{2-17} 
        & $S_{49}$ & & & & \checkmark & & & & & & & & & & &   \\
        \cline{2-17}
        & $S_{63}$ & \checkmark & & & & & & & & & & & & & &   \\ 
        \midrule
        
        \multirow{5}*{\rotatebox[origin=c]{90}{HR}} & 
         $S_{10}$ & \checkmark & & \checkmark & & & & & & & & & & & &    \\
        \cline{2-17}
        & $S_{27}$ & \checkmark & & \checkmark & & & & & & & & & & & &  \\
        \cline{2-17}
        & $S_{28}$ & \checkmark & & \checkmark & & & & & & & & & & & &   \\
        \cline{2-17}
        & $S_{41}$ & \checkmark & & & & & & & & & & & & & &  \\
        \cline{2-17}
        & $S_{43}$ & \checkmark & & \checkmark & & & & & & & & & & & &   \\
        \midrule
        
        \multirow{4}*{\rotatebox[origin=c]{90}{SV}} & $S_{33}$ & & & & & & & \checkmark & & & & & & \checkmark  & &   \\
        \cline{2-17}
        & $S_{34}$ & & & & & & & & & & & & & \checkmark  & &   \\
        \cline{2-17} 
        & $S_{47}$ & & & & & & & & & & & & & \checkmark & &    \\
        \cline{2-17}
        & $S_{48}$ & & & & & & & & & & & & & & & \checkmark  \\
        \midrule
        
        \multirow{1}*{\rotatebox[origin=c]{90}{CL}} & $S_7$ & \checkmark & & & & & & & & & & & & & &   \\
        \midrule
        
        \multirow{1}*{\rotatebox[origin=c]{90}{AA}} &  $S_{50}$ & \checkmark & & & & & & & & & & & & & &   \\
    \hline
\end{longtable}
\end{small}

\subsection{Step 4: learning}
\label{mining-approach}
To extract CTI from textual data, researchers have utilized NLP techniques combined with ML models. Using the methodology described in Section~\ref{coding}, we identify these techniques and the agreement score is $0.48$. In the following ten subsections (\ref{approach:topic} - \ref{approach:report-classification}), we discuss the techniques used by the researchers for training the model for learning to extract CTI. These techniques are presented for each of the CTI extraction purposes, as mentioned in Section~\ref{goal}. Note that a study can use more than one technique and the order of the techniques mentioned in this section is inconsequential. Moreover, the exact implementation may vary from study to study. Table~\ref{tab:extraction_techniques} summarizes the techniques used for each study.  

\subsubsection{CTI-related text classification:}
\label{approach:topic}
In this subsection, the key techniques for CTI-related text classification are given below. 

\begin{itemize}[leftmargin=3mm]
    \item \textbf{Classification (n=10):} Machine learners are trained with examples of a number of classes and then applied for the prediction of the class of unseen data. Natural language-based features such as Term Frequency–Inverse Document Frequency (TF-IDF) \cite{salton1988term} and word embedding~\cite{teller2000speech} are computed from the text segments. Finally, these features are passed to supervised learners, such as support vector machine \cite{cortes1995support}, neural networks \cite{hopfield1982neural}, and naive Bayesian learner~\cite{hand2001idiot} to classify the text segments (e.g., whether the text segments are CTI related or not). This technique is applied for identifying CTI relevant text segments, such as tweets and hacker posts. In $S_3$, $S_{12}$, $S_{14}$, $S_{20}$, $S_{21}$, $S_{26}$, $S_{30}$, $S_{32}$, $S_{38}$, $S_{39}$, the authors utilized these techniques~\footnote{Besides these studies, classification is used in other studies as well, such as with NER or topic modelling. But we did not consider those studies using classification as the technique is already used inherently. For example, to use NER, classification must be used inherently}. 
    
    \item \label{itm:topic-modelling} \textbf{Topic modelling (n=2):} Topic modelling technique is applied for discovering the abstract topics in the text~\cite{topicmodellingwiki}. This technique is taken for generating topic words (such as data breach, denial of service attack) for the CTI-relevant text segments such as tweets, hacker forum posts. Topic models, such as Latent Dirichlet Allocation (LDA) \cite{blei2003latent} and Latent Semantic Analysis (LSA) \cite{landauer1998introduction}, are applied to the text segments to generate topics from the text segments.  In $S_{12}$, $S_{14}$, the authors utilized this technique. 
    
    \item \textbf{Clustering (n=1):} Text segments are clustered based on their similarity through unsupervised learners such as K-means\cite{jain1999data}; similarity computing algorithms such as cosine similarity~\cite{pang2005introduction}; and Jaro Winkler~\cite{bo2019tom} distance algorithms. Clustering techniques are utilized for grouping threat reports, online articles, social media, and forum posts based on their content similarity, such as threat reports discussing similar cyberattacks. In $S_{21}$, the authors utilized this technique. 
    
    \item \textbf{Term weighting (n=1):} Existence of CTI terms is utilized by the researchers to calculate whether the term containing segment denotes an emerging CTI relevant issue. In $S_{31}$, the rate of information gain of CTI relevant terms is used to rank the segments based on CTI relevance or to determine whether the segment contains an emerging topic. 
    
    \item \textbf{Graph embedding (n=1):} Instead of using TF-IDF or word embedding~\cite{teller2000speech}, graph embedding~\cite{goyal2018graph} is also utilized for CTI-relevant text classification. As discussed in \cite{samtani2020proactively}, traditional word embedding or TF-IDF results in high dimensional features which makes the learning inefficient. The authors constructed graph-of-words and graph embedding techniques (such as node2vec\footnote{http://snap.stanford.edu/node2vec/}) to identify the CTI relevant text segments. In $S_{52}$, the authors utilized this technique. 
    
\end{itemize}

\subsubsection{Attack tactics, techniques, and procedures extraction}
The key techniques for extracting attack TTPs in the studies are discussed below.  

\begin{itemize}[leftmargin=3mm]
    \item \textbf{Dependency parsing (n=5):}  The sentences describing cyberattacks contain subjects, verbs, and objects. The subject and object can be an attacker, organization, or tool. The verb denotes the action being performed by attackers or their tools. Utilizing the contextual grammar of English, subjects, verbs, objects, and their dependencies with surrounding terms can be learned. First, the cyberthreat indicators, related noun phrases, and threat actions are annotated. Then using a dependency parser (such as Stanford Typed Dependency Parser\footnote{https://github.com/stanfordnlp/CoreNLP}), the word-word relationship between these terms can be identified and represented as a graph. Then using the dependency graph, the identified keywords can be mapped into TTPs. In $S_2$, $S_{37}$, $S_{55}$, $S_{60}$, $S_{62}$, the authors utilized this technique. 
    
    \item \textbf{Named entity recognition (n=4):} Named entity recognition (NER) refers to the task of classification of entity types mentioned in texts~\cite{nerwiki}. NER can be applied to identify cyberthreat-related techniques and procedures as is done with NER  for extracting IoCs. Text containing TTP-related words are annotated first and then CRF \cite{wallach2004conditional} is used to identify the relevant words' entity types from the CTI candidate text. However, in this technique, establishing the relationship among these identified words is not possible. In $S_5$, $S_{15}$, $S_{37}$, $S_{55}$, the authors utilized this technique.
    
    \item \textbf{Topic modelling (n=2):} 
    Topic modelling can generate a set of probable cyberattack pattern-related topics given a text. Then, using classification, threat reports can be classified to attack patterns. In $S_1$, $S_{35}$, the authors utilized this technique.
    
    \item \textbf{Classification (n=1):}  Threat reports are annotated based on their attack patterns. Then using TF-IDF or word embedding as features, supervised learners can predict the attack pattern for threat reports. In $S_8$, the authors utilized this technique.
    
    \item \textbf{Cybersecurity ontology (n=2):} Unified cybersecurity ontology (UCO), proposed in~\cite{syed2016uco} based on STIX, is used in $S_{15}$, $S_{60}$  to represent the malware, their attack patterns, vulnerabilities, affected systems, and other related entities with a graph-based data structure. 
\end{itemize}

\subsubsection{Cybersecurity-related keywords extraction}
We discuss the key techniques for cybersecurity-related keywords extraction below.  

\begin{itemize}[leftmargin=3mm]
    \item \textbf{Part of speech tagging (n=3):} Part of speech (POS) tagging refers to the mapping of words in a corpus to the corresponding POS such as nouns and verbs~\cite{pos-wiki}. This technique can be used to extract cybersecurity-related noun keywords from CTI relevant text. In $S_9$, $S_{40}$, $S_{51}$, the authors utilized this technique. 
    
    \item \textbf{Named entity recognition (n=3):} NER technique is applied for identifying CTI-relevant words from text segments along with their categories, such as threat types, organization name, organization type, and software names. First, the identification of POS, construction of regular expression patterns, and annotation of entities are performed. Then specific classifiers for the NER task \cite{mohit2014named}, such as conditional random field (CRF) \cite{wallach2004conditional}, is used to learn the patterns of these threat-related entities. In $S_{18}$, $S_{44}$, $S_{51}$, the authors utilized these techniques.
    
    \item \textbf{Topic modelling (n=2):} This technique is applied for generating topic keywords from text (similar to the  Section~\ref{approach:topic}). In $S_{29}$, $S_{61}$, the authors utilized this technique. 
    
    \item \textbf{Term weighting (n=2):} The existence of the CTI-relevant terms is utilized by the researchers to calculate whether the term-containing segment denotes an emerging CTI relevant topic. In $S_{23}$, $S_{53}$, the authors proposed a weight calculation algorithm based on keyword's frequency, security experts' relevance, and the rate of frequency change of the keywords in a given time window to calculate the trendiness of the term. 
    
    \item \textbf{Sentiment analysis (n=1):} Sentiment analysis techniques can be used to analyze the online activities in hacker and darknet forums. In $S_{40}$, the authors computed the sentiment score of posts containing critical infrastructure-related keywords (i.e., banks, hospitals) and cyberthreat related keywords (i.e., malware). Then the authors associated the text segments, keywords, and the sentiment score to identify posts with negative sentiments and thus predict whether certain critical infrastructures are susceptible to cyberattacks. 
    
\end{itemize}

\subsubsection{Cybersecurity-related event identification}
Below, we discuss the key techniques used in the  studies for cyberthreat-related event detection. 

\begin{itemize}[leftmargin=3mm]
    \item \textbf{Classification (n=5):} The event-describing texts are first labelled and then supervised learners are used to train from the examples and predict whether the text is a cybersecurity event or the type of event description in the text. Features, such as distributed topic models, word-word dependency, and existence of named entities, are also used for learning. In $S_{17}$, $S_{22}$, $S_{45}$, $S_{54}$, $S_{64}$,  the authors utilized this approach.
    
    \item \textbf{Clustering (n=3):} The event-describing texts are clustered by similarity of the texts. Cosine similarity, kernel similarity\cite{kate2008dependency, khandpur2017crowdsourcing}, and locality sensitive hashing (LSH) distance\cite{charikar2002similarity} similarity functions are used in $S_4$, $S_{16}$, $S_{56}$, respectively, for clustering. 
    
    \item \textbf{Multitask learning (n=1):} Multi task learning is needed to solve more than one learning task simultaneously while utilizing the overlaps and differences in the learning tasks~\cite{multitaskwiki}. Ji et al. ($S_{36}$) proposed usage of multitask learning in cyberthreat event detection. In this approach, three different sets of features are used by three different learners. These feature sets are organization-specific features (i.e., organization-specific keywords), threat-specific features (i.e., data breach), and generic features (i.e. a common set of words used in both cybersecurity and non-cybersecurity domains). Then multitask learners such as LASSO\cite{zhou2011malsar} is used to learn an optimal model for identifying cyberthreat events.
    
    \item \textbf{Named Entity Recognition (n=4):} Through applying NER, cyberthreat-related keywords and their entity types are identified. The identified keywords are then used as a dictionary for cybersecurity-related words. Cybersecurity events then can be detected based on the existence of cybersecurity keywords in CTI-candidate text and the dictionary. This technique is used by authors in $S_4$, $S_{17}$, $S_{22}$, $S_{54}$.

    \item \textbf{Dependency parsing (n=1):} Dependency parsing is used to identify new words related to specific cybersecurity domain names from CTI-candidate texts. For example, "data breach" being a domain name for events, words such as "data leak" can be found using dependency parsing. This technique is used in $S_{16}$.
    
    \item \textbf{Topic Modelling (n=2):} This technique is applied for generate topic keywords from text which can summarize the content of CTI-candidate text. Then the model is used as a layer for BiLSTM model for reducing feature size. In $S_{17}$, $S_{22}$, the authors utilized this technique.
\end{itemize}

\subsubsection{Indicators of compromise extraction}
To extract indicators from text, the CTI candidate text segments first need to be identified (Section~\ref{approach:topic}). Then the following key techniques are applied.

\begin{itemize}[leftmargin=3mm]
    \item \textbf{Regular expression (n=6):} Cyberthreat indicators. such as malware hash, IP addresses, and software version names. contain specific spelling patterns which can be captured with the use of regular expressions. Based on these spelling patterns (such as OpenIoC\cite{openioc} patterns), indicators can be extracted. In $S_6$, $S_{11}$, $S_{13}$, $S_{42}$, $S_{46}$, $S_{58}$, the authors utilized this technique. 
    
    \item \textbf{Named entity recognition (n=5):} Cyberthreat indicators, such as attack group name, target organization name, vulnerable application name, and malware names do not contain predefined spelling patterns that can be captured by regular expression-based rules and hence, NER technique can be used. From the text segments, researchers annotate the indicator words with their corresponding indicator types. Then conditional random field-based classifiers \cite{wallach2004conditional} (such as Stanford CoreNLP\footnote{https://github.com/stanfordnlp/CoreNLP} CRF) are used to predict the indicator type of words from unseen text segments. In $S_6$, $S_{11}$, $S_{13}$, $S_{57}$, $S_{58}$, the authors utilized this technique. 
    
    \item \textbf{Dependency parsing (n=4):} As the volume of threat-related articles grows over time, new indicators are consistently being mentioned in these articles. Use of regular expression and NER \cite{mohit2014named} may not be able to capture recently-identified indicators. However, the word-word dependency and semantic similarity of word tokens can be utilized in this case. The pattern of dependencies among the POS can be learned, such as subject-verb-object combination. Moreover, semantic similarity based on the contextual features (such as word embedding) of indicator tokens can also be used to predict new indicators. In $S_6$, $S_{11}$, $S_{57}$, $S_{58}$, the authors utilized this technique. 
    
    \item \textbf{Using seed IoCs (n=1):} Many recently-identified indicators coexist with the known indicators in threat-related documents. This intuition is utilized by Zhang et al. ($S_{42}$), where they constructed a set of seed indicators and then using regular expressions, identified new indicators co-existing with these seed indicators mentioned in articles returned by search engines (such as Google). 
    
    \item \textbf{Graph mining (n=1):} Graph mining is a special case of structured data mining where information are extracted from structured or semi-structured data~\cite{graphmining-wiki}. The dependency relation (grammar, POS) of words can be represented as a graph, then the constructed graph can be mined to identify the relations of IoC and context words. This technique is utilized in $S_6$. 
    
    \item \textbf{Clustering (n=1):} In $S_{59}$, the authors used the clustering technique. They clustered all similar malware names using the Jaro Winkler distance algorithm. For each of the malware in the cluster, then they applied K-means~\cite{steinley2006k} clustering with cosine similarity distance on the threat reports, which produces similar threat reports discussing the same malware. Then they combine the malware indicator information from all reports in the same cluster. Note that the threat reports used by the authors are in a semi-structured format (list of key-value pairs).
    
    \item \textbf{Classification (n=1):} A binary classifier is trained from training dataset of potential IoCs, and bag-of-words is used as feature. This technique is used in $S_{42}$.
\end{itemize}

\subsubsection{Cyberthreat alert generation}
The key techniques for threat alert generation in the studies are discussed below. 

\begin{itemize}[leftmargin=3mm]
    \item \textbf{Term weighting (n=3):}   Threat-related terms, such as malware name, attack group name, and specific attack type name (i.e., data breach), are identified, counted, and weighted. Then, based on the keyword frequency and weight, the time window (from when the term is first introduced through the current time/user defined time), and the probability of the keywords appearing in the text, the attack warnings can be generated. In $S_{24}$, $S_{25}$, $S_{49}$, the authors utilized this technique.
    
    \item \textbf{Named entity recognition (n=1):} Through applying NER \cite{mohit2014named}, cyberthreat-related keywords and their entity types can be identified. Then custom rules can be generated based on the word count, probability, and time window selected by the users. In $S_{19}$, the authors utilized this technique.
    
    \item \textbf{Classification (n=1):} In the CTI candidate text, existing threat alert-related texts are labelled and then supervised learners are used to predict any text that can be triggered as a new threat alert. In $S_{63}$, the authors utilized this technique.
    
    \item \textbf{Ontology (n=1):} The relationships among the cybersecurity-related entities are computed using the UCO cybersecurity ontology. Then the ontology is used for applying reasoning over event description for generating alerts. This technique is used in $S_{19}$.
\end{itemize}

\subsubsection{Hacker resource analysis}
The key techniques used in the  studies for analyzing these hacker resources are discussed below.

\begin{itemize}[leftmargin=3mm]
    \item \textbf{Classification (n=5):} Source code, binary programs and executable scripts used by hackers to launch cyberattacks are classified to the corresponding programming/script languages by supervised learners. In $S_{10}$, $S_{27}$, $S_{28}$, $S_{41}$, $S_{43}$, the authors utilized this technique. 
    
    \item \textbf{Topic modelling (n=4):} Topic modelling technique is used to generate cyberthreat related topics from hacker forum posts. In $S_{10}$, $S_{27}$, $S_{28}$, $S_{43}$, the authors utilized this technique. 
\end{itemize}

\subsubsection{Software vulnerability information extraction}
Below, we discuss the key techniques found in the studies for vulnerability-related information extraction.

\begin{itemize}[leftmargin=3mm]
    \item \textbf{Named Entity Recognition (n=1):} Through applying NER, cyberthreat-related keywords and their entity types are mapped to UCO cybersecurity ontology. This technique is used in $S_{33}$.

    \item \textbf{Cybersecurity ontology (n=3):} The vulnerability-related entities are extracted using NER. Then, the relationships among the extracted entities are computed using cybersecurity ontologies, such as UCO. The contextual information of the entities are also retrieved from external resources such as DbPedia\footnote{https://wiki.dbpedia.org/}, MITRE ATT\&CK~\cite{attack} taxonomy, and NVD database. This technique is used in $S_{33}$, $S_{34}$, $S_{47}$. 
    
    \item \textbf{Prediction with historical n-gram (n=1):} This technique is used by Murtaza et al. ($S_{48}$). They predicted the association of applications with any specific type of vulnerability based on prior textual vulnerability description in NVD. They tokenized all NVD description text and computed the n-gram\cite{suen1979n} of all prior vulnerability descriptions. Then, they predicted the association of vulnerability names and application names by the probability computed from the n-grams.
\end{itemize}

\subsubsection{Threat report classification}
\label{approach:report-classification}
Below, we discuss the key techniques found in the studies for threat report classification.

\begin{itemize}[leftmargin=3mm]
    \item \textbf{Classification (n=1):} Threat reports are assigned labels which are the name of the malware discussed in the reports. Then using word embedding as a feature and supervised learners, such as support vector machine, neural networks, K-nearest neighbors \cite{peterson2009k}, the report classification task is trained. In $S_7$, the authors utilized this technique. 
\end{itemize}

\subsubsection{Cyberthreat actor attribution}
Below, we discuss the key techniques for cyberthreat actor attribution. 

\begin{itemize}[leftmargin=3mm]
    \item \textbf{Classification (n=1):} The cyberthreat actors mentioned in the text segments are mapped (one to one, one to many) to the attack patterns mentioned in the reports. Perry et al. ($S_{50}$) manually labelled threat reports with the mentioned threat actors in the threat reports and then used supervised learners to predict threat actors for the unseen threat report. 
\end{itemize}

\subsection{Step 5: exporting to structured format}
\label{sharing-format}
After extraction, CTI could be presented and shared in a structured format. This structured format can be a standard format for sharing CTI such as STIX \cite{stix}, or a general structured format such as Extensible Markup Language (XML) \cite{xml}. Using the methodology described in Section~\ref{coding}, we identify these sharing formats, and the agreement score is $1.00$. We provide more details on the sharing formats and platforms that are used in  studies in the following subsections:

\subsubsection{STIX (n = 2)}
STIX is one of the most commonly used structured language and serialization formats to share CTI in enterprise organizations \cite{shackleford2015s}. STIX information is human and machine readable in JSON and contains domain and relation objects \cite{stix}. Husari et al. ($S_{60}$) mapped the extracted threat actions to techniques, tactics, and kill chain phases and represent CTI in STIX. Ramnani et al. ($S_{55}$) also model the extracted TTPs based on STIX.

\subsubsection{OpenIoC (n = 1)}
OpenIoC is an open source standardized framework for sharing CTI \cite{openioc}. OpenIoC format is based on XML and  is machine-readable. By using the OpenIoC framework, organizations can have access to the latest IoCs shared by other organizations and can communicate with each other \cite{whyopenioc}. Liao et al. ($S_6$) identified the IoCs and corresponding context terms from blog posts and generated an OpenIoC record for each identified IoC. 

\subsubsection{MISP (n = 1)}
MISP \cite{misp} is an open source CTI platform for gathering, storing, and sharing CTI. Using MISP the CTI can be stored in a structured format and can be exported in formats such as STIX, OpenIoC, XML, or CSV. MISP also provides a CTI sharing format based on JSON \cite{misp_dataformat}. Alves et al. ($S_{21}$) used this provided MISP format to generate IoCs that are extracted from the Twitter posts. They chose this format because tweets can have unpredictable content and MISP format is extendable, adaptable, and can be easily converted to other sharing formats such as STIX. 

\subsubsection{Cybersecurity knowledge graph (n = 5)}
A knowledge graph is a set of entity pairs and the relationship between them. Cybersecurity Knowledge Graph (CKG) represents CTI as a knowledge graph \cite{piplai2020creating}. To present CTI in the knowledge graph, researchers use resource description framework (RDF) and unified cybersecurity ontology (UCO). RDF is a standard format and linked to the data representation for data interchange on the Web \cite{RDF}. Unified cybersecurity ontology (UCO) \cite{syed2016uco} is an ontology based on STIX. Piplai et al. ($S_{15}$) described a system to extract information from the security report and represent that in a CKG. They identified the entities (such as malware and tools), the relationships between them, and asserted them in the CKG. They also used UCO to provide cybersecurity domain knowledge in their system's knowledge graph. Neil et al. ($S_{47}$) extracted information on vulnerable packages and dependencies from open source projects and libraries from code repository issues and bug reports. Their extracted CTI is represented in RDF format as a security knowledge graph. Then, they used SPARQL \cite{sparql}, a query language for RDF, to query the obtained CTI. Mittal et al. ($S_{19}$) discover CTI from twitter, represent the gathered CTI using RDF format. They stored RDF representations in a ``Cybersecurity Knowledge Base'' and use SWRL rules \cite{swrl} to generate alerts from this knowledge base. Joshi et al. ($S_{33}$) , and Mulwad et al. ($S_{34}$) also used RDF and created knowledge bases of extracted CTIs.

\subsubsection{General structured format (n = 2)}
Not all researchers used CTI sharing standard formats to represent and share the extracted CTI. For example, Bo et al. ($S_{59}$) proposed a threat operating model that captures the information of cyberthreats gathered from publicly available threat reports and presented them in an XML format \cite{xml}. They used this information to achieve early warning on cyberattacks. Li et al. ($S_{54}$) defined a CTI template to represent CTI in security articles. Their CTI template has two parts including CTI related entities and summarizations of the article. The entities are CVEs, victimized devices, device manufacturers, and impacted locations. These entities and summerizations of each article help their system users to analyze massive open-source data.

\subsection{Step 6: Applications}
\label{use-cases}
In the studies, the authors first extracted CTI from the CTI-candidate text and then demonstrated how they utilized the CTI in application scenarios. In this section, we discuss these applications. 

\subsubsection{Threat landscape (n=6)}
The extracted CTI from the text can give security experts insights of the threat landscape. For example, Liao et al. ($S_6$) the observed the largest number of extracted IoCs are with the type ``PortItem/remoteIP'', which shows the popularity of download-driven phishing and other web-based attacks in the landscape. Macdonald et al. ($S_{40}$) identified potential threats to critical infrastructures. For example, they found a strong relationship between ``DDoS'' and ``bank'' that shows the popularity of financial institutions being targets for DDoS attacks. Husari et al. ($S_{62}$) developed \textit{ActionMiner} to extract threat actions and showed that their results can help to understand the threat landscape. For example, they found ``process'', ``DLL'', ``code'', ``library'', and ``Chrome'', which are the five most related objects to the action ``inject''. This information can also help security experts to plan for mitigating injection attacks. 

Zhao et al. ($S_{58}$) extracted CTI with domain tags, such as industrial control system (ICS) and internet-of-things (IoT). After clustering CTIs based on their domains and analysis of the clusters, they identified insights on different attack types in different domains. For example, they found that the implementation of DDoS attacks varies across multiple domains, and the complexity of the phishing attack depends on the value of the target domain. They also found that IoT-related threats have developed rapidly because of the growing number of IoT devices in recent years. Based on their proposed metrics to quantitatively measure the threat severity from the perspective of security-related social opinion, ICS and governments have experienced threats with higher severity impact than those of other domains. 

Samtani et al. ($S_{52}$) focused on two types of denial of service (DoS) and web application threats that target PHP technology at intervals of three months. Their results showed that the DoS threat landscape is growing more rapidly than the web applications. They found that although new threat types are emerging in both threat categories, the core functions of the threats remain the same over time. This information can guide cybersecurity experts in the prioritization of activities to mitigate threats. Nagai et al. ($S_{61}$) collected IoCs and showed that their approach can help security experts understand attack methods and threat trends in the IoT industry and financial institutions. For example, their results showed that in the IoT industry, the attack methods are being focused on firmware. A relationship between Mirai malware and IoT devices such as routers and printers were found in 2017, which confirms the attack by this malware on IoT devices worldwide at the same period of time.

Extracting CTI from a corpus of articles may also show connections between threats that was never known before. For instance, Liao et al. ($S_6$) clustered the articles in their dataset into 527 clusters based on having at least one IP, email, or domain in common. After analyzing the clusters, they found that the authors of these articles did not realize that the attacks they were documenting were related to other attacks. They also observed that the IoCs reported by a large number of articles disappear quickly. For example, 92\% of IoCs are mentioned on average with 68 articles per month during 0 to 1 month time window before they are stopped.

\subsubsection{Querying CTI (n=4)}
The extracted CTI from the unstructured text can be queried to find information if presented in a queryable platform. For instance, Samtani et al. ($S_{10}$, $S_{27}$) collected malicious attachments and source code from hacker communities, and enabled searching, sorting, and browsing those data through a portal. They also provide a dashboard to show the hacker resource trend over time, key hackers that use those resources, and a list of sources. The information in the dashboard can be filtered by time, by resources, and by a hacker. Neil et al. ($S_{47}$) extracted CTI from open source projects and libraries. They presented software dependencies in a security knowledge graph. Before using a project or a library, developers can query the security knowledge graph and find known vulnerabilities. Piplai et al. ($S_{15}$) built a cybersecurity knowledge graph (CKG) for malware that allows querying the entities of the CKG. To query the CKG, they use SPARQL \cite{sparql} , a query language for RDF. For instance, one possible question to ask is what ‘Tool’ a particular malware uses. 

\subsubsection{Dataset generation (n=4)}
Extracted CTI from the text can contribute as a dataset for use by security researchers and practitioners. In $S_{45}$, the authors developed a new dataset for cyberthreat event identification that is manually annotated and compatible with word embedding based deep learners. Moreover, in $S_{15}$, $S_{33}$, $S_{34}$, the authors stored their extracted CTI in a graph-based data structure that can be consumed by security researchers and other CTI platforms.

\subsubsection{Establishing correlation with attack groups and key hackers (n=3)}
CTI candidate texts often contain information on cybersecurity incidents and associated cyberthreat actors, such as their roles, strategies, and procedures. These information can be used to map the attacks to the responsible cyberthreat actors. For example, the authors in $S_1$ selected 36 cyberthreat actors and an average of 9 CTI documents for each. After extracting TTPs from the documents, they mapped the attack patterns to the responsible cyberthreat actors. Then for an unseen CTI candidate text, their system can predict the cyberthreat actors.  Moreover, in $S_{27}$, $S_{41}$, the authors identified the social networks of cyberthreat actors from CTI candidate texts. Each publication listed identified key cyberthreat actors such as ``KriPpLer'' ,and ``mjrod5'' in $S_{27}$ and ``LinX64'', and ``AsAs'' in $S_{41}$. 

\subsubsection{Malware Protection improvement (n=3)}
Security protection organizations can use the analysis from the extracted CTI to respond quickly. For example, Liao et al. ($S_6$) estimated the time intervals between the first appearance of the IoCs and their adoption by anti-malware tools and web scanners. They observed that, 47\% of the IoCs were detected by anti-virus scanners or IP/URL scanners before they were reported by the technical blogs. For the remaining IoCs, the duration between the first IoC being released and uploaded for a scan is between 0-2 days to more than 12 days. For IPs and domains, the whole process often took more than 12 days. However, malware hashes were often quickly added to anti-virus scanners for scanning, in most cases within 2 days. 

Zhu et al. ($S_{11}$) showed that detection systems (e.g. anti-virus scanners) focus more on blocking network intrusion and removing malicious programs, and they are not capable to detect attacks that use social engineering to download payloads. Hence, a campaign attack can continue for more than one year even after its discovery. This information helps anti-virus vendors to know the weakness of their tools. Williams et al. ($S_{43}$) collect exploit information from hacker forums. They visualized data based on posted exploits and author activity. Data collection was done incrementally, so the information about recent exploits can be helpful for security experts to find new threats. Considering author activity is also valuable to find the most active hacker communities and the exploits they shared.

\subsubsection{Increase awareness of cyberattacks (n=3)}

Extracting CTI from online resources can help security experts to be aware of possible future attacks and to predict and prevent them more effectively. DISCOVER ($S_{24}$) is an early cyberthreat warning system that uses Twitter, cybersecurity blogs, and darkweb forums to generate warnings based on novel terms in these data sources that co-occur with context terms. For example, the NotPetya malware attack went public on 27 June 2017, but DISCOVER generated the first warning related to the malware in February 2017 based on security blogs and in March 2017 based on Twitter data. Neil et al. ($S_{47}$) extracted CTI from open source projects and libraries and present software dependencies in a security knowledge graph. An alert generation system can use this security knowledge graph and generate alerts if a developer can link a library to known vulnerabilities, or if a client installs a vulnerable application. Dion{\'\i}sio et al. ($S_{18}$) showed that their model can identify security-relevant tweets with labelled NER \cite{mohit2014named} from tweets published from 1 to 148 days before the NVD disclosure of a vulnerability. The CVSS\footnote{\url{https://nvd.nist.gov/vuln-metrics/cvss}} severity of tweets ranges from 4.9 to 9.9 that show the importance of alert generation of the model.

\subsubsection{Discovery of zero-day exploits (n=1)}
One of the applications of extracting CTI is to discover zero-day exploits. Detection of these exploits at an earlier stage can help organizations protect their system or minimize the damage caused by the attack \cite{nunes2016darknet}. Nunes et al. ($S_{20}$) detected 16 zero-day exploits from darknet marketplace data in a 4-week period. 

\subsubsection{Cross-site connection between multiple CTI sources (n=1)}
If CTI extraction is done on more than one source, the extracted data can reveal cross-site connections between sources. For example, Nunes et al. ($S_{20}$) used darknet marketplaces and hacker forums to collect CTI. They created a connected graph using the ``usernames'' used in two domains. They found individuals selling products related to malicious hacking in marketplaces and hacking forums simultaneously. This information is helpful to determine the social groups of the domain.

\subsubsection{CTI relationship analysis (n=1)}
From the extracted CTI, relationship analysis can be performed among them such as the association between threats, techniques, tools, mitigation. For example, Piplai et al. ($S_{15}$) build a cybersecurity knowledge graph for malware where they included details such as malware's campaign, used tools, and targeted software in the knowledge graph. The graph can be used to compare malware and cluster similar malware. 

\subsubsection{APT technique analysis (n = 1)}
Analyzing APT~\cite{chen2014study}-related technique trends can result in valuable insights, such as the trends of the attack techniques used. As an example, Niakanlahiji et al. ($S_2$), analyse the trend for the 14 most-mentioned techniques in their data. They found that from 2013 to 2016, exploiting browsers and using malicious scripts are the most used techniques. In 2017, the use of PowerShell became one of the top techniques; using malicious scripts remained one of the top techniques; but exploiting browsers became insignificant. This type of information can inform security experts to prioritize mitigation strategies. In the same study, researchers analyzed the relationship between together-mentioned techniques and calculated the strength of these relationships. For instance, they observe a strong relationship between obfuscation and using scripts in their APT reports and show that APTs commonly use obfuscation to protect their scripts.

\subsubsection{Malware profiling (n=1)}
Malware can be profiled based on CTI candidate texts. For example, Bo et al. ($S_{59}$) extracted the malware information from multiple threat reports and constructed the malware profile in an XML format. The malware profile contains static attributes, such as operating system, platform, and CVE; and dynamic attributes, such as hotness (the degree to which the malware is attacking lately) and hack interest (the degree to which the malware is attacking frequently). For example, in the profile of the malware W32.Kwbot.Worm operating system is ``Windows CE; UNIX'', CVE is ``CVE-2012-0158'', hotness is 149, and hack interest is 9913. This information can help organizations to check their IT environment and generate early warnings. 

\section{Threats to validity}
The search process of finding the relevant publications may not be comprehensive as we use six scholar databases as sources. However, other scholar databases may contain more studies. The process of searching, applying filtering criteria, and coding are subjective. The findings in our article are based on 64 selected studies. Moreover, the generality of our findings may be limiting as the process of searching and documenting the findings from each study is a subjective process.

\section{Challenges}
\label{challenge}
In this section, we discuss four technical challenges for future researchers in the CTI extraction domain. 

\begin{enumerate}[leftmargin=5mm, label=\alph*)]
    \item \textbf{Need for clean, labelled, and published datasets:} Collecting and cleaning data are the first step done by researchers and can be time consuming and complex. If cleaned and labelled data is available, then other researchers can use the dataset with no further effort. In addition, in supervised tasks, such as classification, labelled data is needed. Even if the raw data is available, the labeling should be done by researchers and the process takes significant time and effort. In most of the studies, sources are mentioned for collecting textual data, such as Fireeye \cite{fireeye} or Twitter, but the collected and labeled data is not shared for future use by others. One possible reason could be the changing nature of attack patterns. Collected data in a period may not be valuable in the future for CTI extraction. However, creating and publishing datasets, especially labeled datasets, should be considered to ease the potential of extending the work by other cybersecurity researchers in the future, for training data, and to enable comparison between analysis techniques.
    
    \item \textbf{Handcrafting ground truth:} With the lack of labeled data, a large number of analyzed studies perform manual annotation or labeling. This task is better to be done by more than one person and takes significant time and effort. For example, Behzadan et al. ($S_{14}$) manually annotated 21,000 cyber-related tweets. They publish their dataset for future usage by researchers. Husari et al. ($S_{60}$) propose and evaluate a threat action ontology. They analyze threat actions of attack techniques and patterns described in MITRE ATT\&CK \cite{attack} and CAPEC\cite{capec} in a top-down approach. Also, to evaluate their ontology, they manually extract 1,512 threat actions from 80 randomly-selected threat reports. Thus, handcrafting ground truth for training and test cases might pose challenges for further CTI extraction research. Therefore, research effort should be focused on preparing ground truth for CTI extractions.

    \item \textbf{Mitigating class imbalance issues:} A challenge researchers face is having imbalanced classes because of the nature of the collected data. For instance, from 1 million news articles, just 500 articles may be CTI related. Imbalance classes can affect the quality of the training of the model and the accuracy of testing the model. In the studies, some of the researchers observe class imbalance and follow an approach to solve the problem. Behzadan et al. ($S_{14}$) use weighted classes of an appropriate ratio to train their classifier, Le et al. ($S_{38}$) use novelty detection approaches which need samples of one class to train the model \cite{frasconi2021two}. Queiroz et al. ($S_{26}$) use random oversampling \cite{chawla2009data} to increase the number of instances of positive classes in their datasets. Mitigating class imbalance issues in dataset can be given priority in the future CTI extraction research.

    \item \textbf{Need for baseline and source code of the proposed approaches for comparison:} Publishing the code repository or implemented approach in the studies we analyzed is rare. One way to evaluate the results of the proposed approach is to compare the results with the available baseline and the results of other approaches in the area of research. Because of not publishing the data and the analysis code for the approach, most of the studies are not replicable and comparable. Therefore, the comparison between existing approaches and the proposed approach in each study is very hard or even impossible. Several studies provide links to their application or code repositories, but the links are not available anymore, such as \url{https://github.com/qclassified/cici/} and \url{https://ioc-chainsmith.org/}. Ghazi et al. ($S_5$) publish the annotation tool they designed and use to train a NER \cite{mohit2014named} model which tags documents to extract CTI concepts, such as Actor and TTPs \cite{g4ti}. Dion{\'\i}sio et al. ($S_{18}$) published their code for cyberthreat detection from Twitter on a GitHub repository\cite{relevanttweets}. They provide pretrained models and evaluation scripts to evaluate them. Users have the ability to train the models themselves as well. Niakanlahiji et al. ($S_{46}$) published their code for IoCMiner, an automated approach to extract IoC from Twitter on GitHub \cite{iocminercode}. Their repository includes raw and labeled data and implemented models. Samtani et al. ($S_{52}$) published their code for proactively identifying emerging hacker threats in the dark web on GitHub \cite{dgefcode}. They propose a Diachronic Graph Embedding Framework (D-GEF) to generate word embeddings in an unsupervised approach.
    
\end{enumerate}

\section{Future research direction}
\label{future}
In this section, we discuss potential future research directions.

\begin{enumerate}[leftmargin=5mm, label=\alph*)]
    \item \textbf{More focus on the actionability of the extracted CTI:} The extracted CTI can be called actionable if the CTI is \textit{relevant} and \textit{trustworthy} to the operations of IT organization, provide \textit{complete} and \textit{accurate} information, and can be \textit{ingested} to other CTI sharing platforms~\cite{conrad1980there, pawlinski2014actionable, pawlinski2015actionable}. More research endeavor should be given to extract the CTI to be actionable enough for the consuming organizations. Moreover, through the usage of the actionable CTI, research focus should also be given to mitigate risks, optimize security practice, and establish correlation between the extracted CTI.   
    
    \item \textbf{Adapting to the changes in attack strategy:} With the passage of time, attackers continue to change their tactics and techniques to bypass and evade defense mechanisms. CTI extracted from current CTI candidate texts may be obsolete or irrelevant in the future. Hence, focus should be given on how researchers can utilize the insight gained from already-extracted CTI to establish a correlation to CTI which will be extracted in the future. 
    
    \item \textbf{Extracting CTI from large, multiple dataset:} In an academic environment, researchers usually work on single dataset and report the observation~\cite{Samtani2020} which results in duplicated efforts from researchers and weak correlation among the extracted CTI. Hence, focus should be given on how cybersecurity researchers can aggregate the information gained extracted CTI from large and multiple dataset, find the relationship among these information, and instantiate actionable knowledge for relevant organizations from this collection of information. Moreover, extracted CTI should be checked for quality issues such as false alarms, and consistency. 
    
    \item \textbf{Prioritization and automated decision making: } CTI candidate texts provide raw data for extracting potential CTI. However, focus should be given on how cybersecurity practitioners can prioritize the proactive action. Moreover, researchers can explore how extracted CTI can be followed by automated decision making based on the extracted intelligence. 
    
    \item \textbf{Exploring cybersecurity language model:} Cybersecurity specific NER model and word embeddings (i.e., sec2vec\footnote{https://github.com/0xyd/sec2vec}, Harvard NER corpus\footnote{https://dataverse.harvard.edu/dataset.xhtml?persistentId=doi:10.7910/DVN/1TCFII}) can also be used instead of stock pretrained models (such as word2vec~\cite{mikolov2013efficient}, glove\footnote{https://nlp.stanford.edu/projects/glove/}) to gain better performance in learning tasks.
    
\end{enumerate}

\section{Conclusion}
\label{discussion}
CTI can be extracted from unstructured texts on cyberthreat-related topics written by cybersecurity researchers, organizations, and vendors. In this survey, we identify 64 relevant studies from 6 scholar databases. We categorize the CTI extraction purposes, propose a CTI extraction pipeline, and identify the data sources,  techniques, and CTI sharing formats utilized in the context of the proposed pipeline. Our work finds ten types of extraction purposes where CTI text classification, attack pattern extraction, and cybersecurity keywords extraction have given greater attention. We also identify seven types of textual sources for CTI extraction where hacker forums, threat reports, social media posts, and online news articles are largely utilized. We also observe that natural language processing and machine learning based techniques, such as supervised classification, named entity recognition, topic modelling, and dependency parsing are the primary techniques used for CTI extraction. Finally, we conclude with a set of technical challenges observed in the studies and future research directions in the CTI extraction domain. Prospective cybersecurity researchers can benefit from our work, such as instantiating their own CTI extraction pipeline based on their extraction purposes, identifying relevant data sources, and selecting appropriate techniques for mining CTI. Overall, the work provides researchers options for making design decisions for their own CTI extraction method from natural language artifacts.  

\section*{Acknowledgement}
\label{sec:acknowledgment}
This work is supported by the NSA Science of Security Lablet and the NSA Laboratory for Analytic Sciences.


\bibliographystyle{ACM-Reference-Format}
\bibliography{reference}

\section{Appendix}
\label{appendix}

\begin{small}

\begin{longtable}{lp{13cm}}
    \caption{Selected studies} 
    \label{tab:selected-studies} \\ \hline
        \textbf{Id} & \textbf{Publication}  \\
        
        $S_1$
        & Noor, Umara, Zahid Anwar, Tehmina Amjad, and Kim-Kwang Raymond Choo. "A machine learning-based FinTech cyberthreat attribution framework using high-level indicators of compromise." Future Generation Computer Systems 96 (2019): 227-242.  \\
        
        $S_2$
        & Niakanlahiji, Amirreza, Jinpeng Wei, and Bei-Tseng Chu. "A natural language processing based trend analysis of advanced persistent threat techniques." In 2018 IEEE International Conference on Big Data (Big Data), pp. 2995-3000. IEEE, 2018.   \\
        
        $S_3$
        & Shin, Han-Sub, Hyuk-Yoon Kwon, and Seung-Jin Ryu. "A new text classification model based on contrastive word embedding for detecting cybersecurity intelligence in twitter." Electronics 9, no. 9 (2020): 1527.  \\
        
        $S_4$
        & Bose, Avishek, Vahid Behzadan, Carlos Aguirre, and William H. Hsu. "A novel approach for detection and ranking of trendy and emerging cyber threat events in twitter streams." In 2019 IEEE/ACM International Conference on Advances in Social Networks Analysis and Mining (ASONAM), pp. 871-878. IEEE, 2019.   \\
        
        $S_5$
        & Ghazi, Yumna, Zahid Anwar, Rafia Mumtaz, Shahzad Saleem, and Ali Tahir. "A supervised machine learning based approach for automatically extracting high-level threat intelligence from unstructured sources." In 2018 International Conference on Frontiers of Information Technology (FIT), pp. 129-134. IEEE, 2018.   \\
        
        $S_6$
        & Liao, Xiaojing, Kan Yuan, XiaoFeng Wang, Zhou Li, Luyi Xing, and Raheem Beyah. "Acing the ioc game: Toward automatic discovery and analysis of open-source cyber threat intelligence." In Proceedings of the 2016 ACM SIGSAC Conference on Computer and Communications Security, pp. 755-766. 2016.   \\
        
        $S_7$
        & Yang, Wenzhuo, and Kwok-Yan Lam. "Automated cyber threat intelligence reports classification for early warning of cyber attacks in next generation soc." In International Conference on Information and Communications Security, pp. 145-164. Springer, Cham, 2019. \\
        
        $S_8$
        & Ayoade, Gbadebo, Swarup Chandra, Latifur Khan, Kevin Hamlen, and Bhavani Thuraisingham. "Automated threat report classification over multi-source data." In 2018 IEEE 4th International Conference on Collaboration and Internet Computing (CIC), pp. 236-245. IEEE, 2018. \\
        
        $S_9$
        & Wang, Tianyi, and Kam Pui Chow. "Automatic Tagging of Cyber Threat Intelligence Unstructured Data using Semantics Extraction." In 2019 IEEE International Conference on Intelligence and Security Informatics (ISI), pp. 197-199. IEEE, 2019. \\
        
        $S_{10}$
        & Samtani, Sagar, Kory Chinn, Cathy Larson, and Hsinchun Chen. "Azsecure hacker assets portal: Cyber threat intelligence and malware analysis." In 2016 IEEE conference on intelligence and security informatics (ISI), pp. 19-24. Ieee, 2016.  \\
        
        $S_{11}$
        & Zhu, Ziyun, and Tudor Dumitras. "Chainsmith: Automatically learning the semantics of malicious campaigns by mining threat intelligence reports." In 2018 IEEE European Symposium on Security and Privacy (EuroS\&P), pp. 458-472. IEEE, 2018.  \\
        
        $S_{12}$
        & Deliu, Isuf, Carl Leichter, and Katrin Franke. "Collecting cyber threat intelligence from hacker forums via a two-stage, hybrid process using support vector machines and latent dirichlet allocation." In 2018 IEEE International Conference on Big Data (Big Data), pp. 5008-5013. IEEE, 2018. \\
        
        $S_{13}$
        & Long, Zi, Lianzhi Tan, Shengping Zhou, Chaoyang He, and Xin Liu. "Collecting indicators of compromise from unstructured text of cybersecurity articles using neural-based sequence labelling." In 2019 International Joint Conference on Neural Networks (IJCNN), pp. 1-8. IEEE, 2019. \\
        
        $S_{14}$
        & Behzadan, Vahid, Carlos Aguirre, Avishek Bose, and William Hsu. "Corpus and deep learning classifier for collection of cyber threat indicators in twitter stream." In 2018 IEEE International Conference on Big Data (Big Data), pp. 5002-5007. IEEE, 2018. \\
        
        $S_{15}$
        & Piplai, Aritran, Sudip Mittal, Anupam Joshi, Tim Finin, James Holt, and Richard Zak. "Creating cybersecurity knowledge graphs from malware after action reports." IEEE Access 8 (2020): 211691-211703. \\
        
        $S_{16}$
        & Khandpur, Rupinder Paul, Taoran Ji, Steve Jan, Gang Wang, Chang-Tien Lu, and Naren Ramakrishnan. "Crowdsourcing cybersecurity: Cyber attack detection using social media." In Proceedings of the 2017 ACM on Conference on Information and Knowledge Management, pp. 1049-1057. 2017. \\
        
        $S_{17}$
        & Abdullah, Mohamad Syahir, Anazida Zainal, Mohd Aizaini Maarof, and Mohamad Nizam Kassim. "Cyber-attack features for detecting cyber threat incidents from online news." In 2018 Cyber Resilience Conference (CRC), pp. 1-4. IEEE, 2018. \\
        
        $S_{18}$
        & Dionísio, Nuno, Fernando Alves, Pedro M. Ferreira, and Alysson Bessani. "Cyberthreat detection from twitter using deep neural networks." In 2019 International Joint Conference on Neural Networks (IJCNN), pp. 1-8. IEEE, 2019. \\
        
        $S_{19}$
        & Mittal, Sudip, Prajit Kumar Das, Varish Mulwad, Anupam Joshi, and Tim Finin. "Cybertwitter: Using twitter to generate alerts for cybersecurity threats and vulnerabilities." In 2016 IEEE/ACM International Conference on Advances in Social Networks Analysis and Mining (ASONAM), pp. 860-867. IEEE, 2016. \\
        
        $S_{20}$
        & Nunes, Eric, Ahmad Diab, Andrew Gunn, Ericsson Marin, Vineet Mishra, Vivin Paliath, John Robertson, Jana Shakarian, Amanda Thart, and Paulo Shakarian. "Darknet and deepnet mining for proactive cybersecurity threat intelligence." In 2016 IEEE Conference on Intelligence and Security Informatics (ISI), pp. 7-12. IEEE, 2016. \\
        
        $S_{21}$
        & Alves, Fernando, Pedro Miguel Ferreira, and Alysson Bessani. "Design of a classification model for a twitter-based streaming threat monitor." In 2019 49th Annual IEEE/IFIP International Conference on Dependable Systems and Networks Workshops (DSN-W), pp. 9-14. IEEE, 2019.  \\
        
        $S_{22}$
        & Fang, Yong, Jian Gao, Zhonglin Liu, and Cheng Huang. "Detecting cyber threat event from twitter using IDCNN and BILSTM." Applied Sciences 10, no. 17 (2020): 5922. \\
        
        $S_{23}$
        & Hughes, Jack, Seth Aycock, Andrew Caines, Paula Buttery, and Alice Hutchings. "Detecting Trending Terms in Cybersecurity Forum Discussions." In Proceedings of the Sixth Workshop on Noisy User-generated Text (W-NUT 2020), pp. 107-115. 2020.  \\
        
        $S_{24}$
        & Sapienza, Anna, Sindhu Kiranmai Ernala, Alessandro Bessi, Kristina Lerman, and Emilio Ferrara. "Discover: Mining online chatter for emerging cyber threats." In Companion Proceedings of the The Web Conference 2018, pp. 983-990. 2018. \\
        
        $S_{25}$
        & Sapienza, Anna, Alessandro Bessi, Saranya Damodaran, Paulo Shakarian, Kristina Lerman, and Emilio Ferrara. "Early warnings of cyber threats in online discussions." In 2017 IEEE International Conference on Data Mining Workshops (ICDMW), pp. 667-674. IEEE, 2017. \\
        
        $S_{26}$
        & Queiroz, Andrei Lima, Susan Mckeever, and Brian Keegan. "Eavesdropping hackers: Detecting software vulnerability communication on social media using text mining." In The Fourth International Conference on Cyber-Technologies and Cyber-Systems, pp. 41-48. 2019. \\
        
        $S_{27}$
        & Samtani, Sagar, Ryan Chinn, Hsinchun Chen, and Jay F. Nunamaker Jr. "Exploring emerging hacker assets and key hackers for proactive cyber threat intelligence." Journal of Management Information Systems 34, no. 4 (2017): 1023-1053. \\
        
        $S_{28}$
        & Samtani, Sagar, Ryan Chinn, and Hsinchun Chen. "Exploring hacker assets in underground forums." In 2015 IEEE international conference on intelligence and security informatics (ISI), pp. 31-36. IEEE, 2015. \\
        
        $S_{29}$
        & Al-Ramahi, Mohammad, Izzat Alsmadi, and Joshua Davenport. "Exploring hackers assets: topics of interest as indicators of compromise." In Proceedings of the 7th Symposium on Hot Topics in the Science of Security, pp. 1-4. 2020. \\
        
        $S_{30}$
        & Kadoguchi, Masashi, Shota Hayashi, Masaki Hashimoto, and Akira Otsuka. "Exploring the dark web for cyber threat intelligence using machine leaning." In 2019 IEEE International Conference on Intelligence and Security Informatics (ISI), pp. 200-202. IEEE, 2019. \\
        
        $S_{31}$
        & Benjamin, Victor, Weifeng Li, Thomas Holt, and Hsinchun Chen. "Exploring threats and vulnerabilities in hacker web: Forums, IRC and carding shops." In 2015 IEEE international conference on intelligence and security informatics (ISI), pp. 85-90. IEEE, 2015.  \\
        
        $S_{32}$
        & Deliu, Isuf, Carl Leichter, and Katrin Franke. "Extracting cyber threat intelligence from hacker forums: Support vector machines versus convolutional neural networks." In 2017 IEEE International Conference on Big Data (Big Data), pp. 3648-3656. IEEE, 2017.  \\
        
        $S_{33}$
        & Joshi, Arnav, Ravendar Lal, Tim Finin, and Anupam Joshi. "Extracting cybersecurity related linked data from text." In 2013 IEEE Seventh International Conference on Semantic Computing, pp. 252-259. IEEE, 2013. \\
        
        $S_{34}$
        & Mulwad, Varish, Wenjia Li, Anupam Joshi, Tim Finin, and Krishnamurthy Viswanathan. "Extracting information about security vulnerabilities from web text." In 2011 IEEE/WIC/ACM International Conferences on Web Intelligence and Intelligent Agent Technology, vol. 3, pp. 257-260. IEEE, 2011. \\
        
        $S_{35}$
        & Li, Mengming, Rongfeng Zheng, Liang Liu, and Pin Yang. "Extraction of Threat Actions from Threat-related Articles using Multi-Label Machine Learning Classification Method." In 2019 2nd International Conference on Safety Produce Informatization (IICSPI), pp. 428-431. IEEE, 2019. \\
        
        $S_{36}$
        & Ji, Taoran, Xuchao Zhang, Nathan Self, Kaiqun Fu, Chang-Tien Lu, and Naren Ramakrishnan. "Feature driven learning framework for cybersecurity event detection." In Proceedings of the 2019 IEEE/ACM International Conference on Advances in Social Networks Analysis and Mining, pp. 196-203. 2019. \\
        
        $S_{37}$
        & Zhu, Ziyun, and Tudor Dumitraş. "Featuresmith: Automatically engineering features for malware detection by mining the security literature." In Proceedings of the 2016 ACM SIGSAC Conference on Computer and Communications Security, pp. 767-778. 2016.  \\
        
        $S_{38}$
        & Le, Ba Dung, Guanhua Wang, Mehwish Nasim, and Ali Babar. "Gathering cyber threat intelligence from Twitter using novelty classification." arXiv preprint arXiv:1907.01755 (2019). \\
        
        $S_{39}$
        & Gautam, Apurv Singh, Yamini Gahlot, and Pooja Kamat. "Hacker Forum Exploit and Classification for Proactive Cyber Threat Intelligence." In International Conference on Inventive Computation Technologies, pp. 279-285. Springer, Cham, 2019. \\
        
        $S_{40}$
        & Macdonald, Mitch, Richard Frank, Joseph Mei, and Bryan Monk. "Identifying digital threats in a hacker web forum." In Proceedings of the 2015 IEEE/ACM international conference on advances in social networks analysis and mining 2015, pp. 926-933. 2015. \\
        
        $S_{41}$
        & Grisham, John, Sagar Samtani, Mark Patton, and Hsinchun Chen. "Identifying mobile malware and key threat actors in online hacker forums for proactive cyber threat intelligence." In 2017 IEEE International Conference on Intelligence and Security Informatics (ISI), pp. 13-18. IEEE, 2017.\\
        
        $S_{42}$
        & Zhang, Panpan, Jing Ya, Tingwen Liu, Quangang Li, Jinqiao Shi, and Zhaojun Gu. "iMCircle: Automatic Mining of Indicators of Compromise from the Web." In 2019 IEEE Symposium on Computers and Communications (ISCC), pp. 1-6. IEEE, 2019.\\
        
        $S_{43}$
        & Williams, Ryan, Sagar Samtani, Mark Patton, and Hsinchun Chen. "Incremental hacker forum exploit collection and classification for proactive cyber threat intelligence: An exploratory study." In 2018 IEEE International Conference on Intelligence and Security Informatics (ISI), pp. 94-99. IEEE, 2018.  \\
        
        $S_{44}$
        & Gasmi, Houssem, Jannik Laval, and Abdelaziz Bouras. "Information extraction of cybersecurity concepts: An LSTM approach." Applied Sciences 9, no. 19 (2019): 3945. \\
        
        $S_{45}$
        & Trong, Hieu Man Duc, Duc Trong Le, Amir Pouran Ben Veyseh, Thuat Nguyen, and Thien Huu Nguyen. "Introducing a New Dataset for Event Detection in Cybersecurity Texts." In Proceedings of the 2020 Conference on Empirical Methods in Natural Language Processing (EMNLP), pp. 5381-5390. 2020. \\
        
        $S_{46}$
        & Niakanlahiji, Amirreza, Lida Safarnejad, Reginald Harper, and Bei-Tseng Chu. "Iocminer: Automatic extraction of indicators of compromise from twitter." In 2019 IEEE International Conference on Big Data (Big Data), pp. 4747-4754. IEEE, 2019.  \\
        
        $S_{47}$
        & Neil, Lorenzo, Sudip Mittal, and Anupam Joshi. "Mining threat intelligence about open-source projects and libraries from code repository issues and bug reports." In 2018 IEEE International Conference on Intelligence and Security Informatics (ISI), pp. 7-12. IEEE, 2018.  \\
        
        $S_{48}$
        & Murtaza, Syed Shariyar, Wael Khreich, Abdelwahab Hamou-Lhadj, and Ayse Basar Bener. "Mining trends and patterns of software vulnerabilities." Journal of Systems and Software 117 (2016): 218-228.\\
        
        $S_{49}$
        & Dong, Fangzhou, Shaoxian Yuan, Haoran Ou, and Liang Liu. "New cyber threat discovery from darknet marketplaces." In 2018 IEEE Conference on Big Data and Analytics (ICBDA), pp. 62-67. IEEE, 2018.  \\
        
        $S_{50}$
        & Perry, Lior, Bracha Shapira, and Rami Puzis. "NO-DOUBT: Attack attribution based on threat intelligence reports." In 2019 IEEE International Conference on Intelligence and Security Informatics (ISI), pp. 80-85. IEEE, 2019. \\
        
        $S_{51}$
        & Li, Dong, Xiao Zhou, and Ao Xue. "Open source threat intelligence discovery based on topic detection." In 2020 29th International Conference on Computer Communications and Networks (ICCCN), pp. 1-4. IEEE, 2020.  \\
        
        $S_{52}$
        & Samtani, Sagar, Hongyi Zhu, and Hsinchun Chen. "Proactively Identifying Emerging Hacker Threats from the Dark Web: A Diachronic Graph Embedding Framework (D-GEF)." ACM Transactions on Privacy and Security (TOPS) 23, no. 4 (2020): 1-33.  \\
        
        $S_{53}$
        & Lee, Kuo-Chan, Chih-Hung Hsieh, Li-Jia Wei, Ching-Hao Mao, Jyun-Han Dai, and Yu-Ting Kuang. "Sec-Buzzer: cyber security emerging topic mining with open threat intelligence retrieval and timeline event annotation." Soft Computing 21, no. 11 (2017): 2883-2896.  \\
        
        $S_{54}$
        & Li, Ke, Hui Wen, Hong Li, Hongsong Zhu, and Limin Sun. "Security OSIF: Toward automatic discovery and analysis of event based cyber threat intelligence." In 2018 IEEE SmartWorld, Ubiquitous Intelligence \& Computing, Advanced \& Trusted Computing, Scalable Computing \& Communications, Cloud \& Big Data Computing, Internet of People and Smart City Innovation (SmartWorld/SCALCOM/UIC/ATC/CBDCom/IOP/SCI), pp. 741-747. IEEE, 2018. \\
        
        $S_{55}$
        & Ramnani, Roshni R., Karthik Shivaram, and Shubhashis Sengupta. "Semi-automated information extraction from unstructured threat advisories." In Proceedings of the 10th Innovations in Software Engineering Conference, pp. 181-187. 2017. \\
        
        $S_{56}$
        & Le Sceller, Quentin, ElMouatez Billah Karbab, Mourad Debbabi, and Farkhund Iqbal. "Sonar: Automatic detection of cyber security events over the twitter stream." In Proceedings of the 12th International Conference on Availability, Reliability and Security, pp. 1-11. 2017. \\
        
        $S_{57}$
        & Kim, Nakhyun, Minseok Kim, Seulgi Lee, Hyeisun Cho, Byung-ik Kim, Jun-hyung Park, and MoonSeog Jun. "Study of Natural Language Processing for Collecting Cyber Threat Intelligence Using SyntaxNet." In International Symposium of Information and Internet Technology, pp. 10-18. Springer, Cham, 2018. \\
        
        $S_{58}$
        & Zhao, Jun, Qiben Yan, Jianxin Li, Minglai Shao, Zuti He, and Bo Li. "TIMiner: Automatically extracting and analyzing categorized cyber threat intelligence from social data." Computers \& Security 95 (2020): 101867.\\
        
        $S_{59}$
        & Bo, Tao, Yue Chen, Can Wang, Yunwei Zhao, Kwok-Yan Lam, Chi-Hung Chi, and Hui Tian. "TOM: A Threat Operating Model for Early Warning of Cyber Security Threats." In International Conference on Advanced Data Mining and Applications, pp. 696-711. Springer, Cham, 2019. \\
        
        $S_{60}$
        & Husari, Ghaith, Ehab Al-Shaer, Mohiuddin Ahmed, Bill Chu, and Xi Niu. "Ttpdrill: Automatic and accurate extraction of threat actions from unstructured text of cti sources." In Proceedings of the 33rd Annual Computer Security Applications Conference, pp. 103-115. 2017. \\
        
        $S_{61}$
        & Nagai, Tatsuya, Makoto Takita, Keisuke Furumoto, Yoshiaki Shiraishi, Kelin Xia, Yasuhiro Takano, Masami Mohri, and Masakatu Morii. "Understanding Attack Trends from Security Blog Posts Using Guided-topic Model." Journal of Information Processing 27 (2019): 802-809. \\
        
        $S_{62}$
        & Husari, Ghaith, Xi Niu, Bill Chu, and Ehab Al-Shaer. "Using entropy and mutual information to extract threat actions from cyber threat intelligence." In 2018 IEEE International Conference on Intelligence and Security Informatics (ISI), pp. 1-6. IEEE, 2018. \\
        
        $S_{63}$
        & Sabottke, Carl, Octavian Suciu, and Tudor Dumitraș. "Vulnerability disclosure in the age of social media: Exploiting twitter for predicting real-world exploits." In 24th {USENIX} Security Symposium ({USENIX} Security 15), pp. 1041-1056. 2015. \\
        
        $S_{64}$
        & Ritter, Alan, Evan Wright, William Casey, and Tom Mitchell. "Weakly supervised extraction of computer security events from twitter." In Proceedings of the 24th International Conference on World Wide Web, pp. 896-905. 2015. \\
    \hline
\end{longtable}

\end{small}

\end{document}